\documentclass[Journal,twocolumn]{IEEEtran}
\IEEEoverridecommandlockouts
\usepackage{color}
\usepackage{graphicx}
\usepackage{amsmath}
\usepackage{amssymb}
\usepackage{algorithm}
\usepackage{algorithmic}
\usepackage{amsmath}
\usepackage{multirow}
\usepackage{booktabs}
\usepackage{array}
\usepackage{amsthm}
\usepackage{stfloats}
\usepackage{caption}
\usepackage{subfigure}
\usepackage{bm}
\usepackage{booktabs}
\usepackage{setspace}
\usepackage{url}

\newcommand{\be}{\begin{equation}}
\newcommand{\ee}{\end{equation}}
\newcommand{\bea}{\begin{eqnarray}}
\newcommand{\eea}{\end{eqnarray}}
\newcommand{\ba}{\begin{array}}
\newcommand{\ea}{\end{array}}
\newcommand{\nid}{\noindent}
\newcommand{\non}{\nonumber}

\allowdisplaybreaks[4]

\title{Intelligent Reflecting Surface based Passive Information Transmission: A Symbol-Level Precoding Approach
\thanks{R. Liu and M. Li are with the School of Information and Communication Engineering, Dalian University of Technology, Dalian 116024, China (e-mail: liurang@mail.dlut.edu.cn; mli@dlut.edu.cn).}
\thanks{Q. Liu is with the School of Computer Science and Technology, Dalian University of Technology, Dalian 116024, China (e-mail: qianliu@dlut.edu.cn).}
\thanks{A. L. Swindlehurst is with the Center for Pervasive Communications and Computing, University of California, Irvine, CA 92697, USA (e-mail: swindle@uci.edu).}
\thanks{Q. Wu is with the State Key Laboratory of Internet of Things for Smart City and Department of Electrical and Computer Engineering, University of Macau, Macau, 999078, China (email: qingqingwu@um.edu.mo).}
}

\author{Rang Liu,~\IEEEmembership{Graduate Student Member,~IEEE,}
        Ming Li,~\IEEEmembership{Senior Member,~IEEE,}
        Qian Liu,~\IEEEmembership{Member,~IEEE,} \\
        A. Lee Swindlehurst,~\IEEEmembership{Fellow,~IEEE,}
        and Qingqing Wu,~\IEEEmembership{Member,~IEEE}}

\begin{document}

\maketitle
\pagestyle{empty}
\thispagestyle{empty}

\begin{abstract}
Intelligent reflecting surfaces (IRS) have been proposed as a  revolutionary technology owing to its capability of adaptively reconfiguring the propagation environment in a cost-effective and hardware-efficient fashion.
While the application of IRS as a passive reflector to enhance the performance of wireless communications has been widely investigated in the literature, using IRS as a passive transmitter recently is emerging as a new concept and attracting steadily growing interest.
In this paper, we propose two novel IRS-based passive information transmission systems using advanced symbol-level precoding.
One is a standalone passive information transmission system, where the IRS operates as a passive transmitter serving multiple receivers by adjusting its elements to reflect unmodulated carrier signals.
The other is a joint passive reflection and information transmission system, where the IRS not only enhances transmissions for multiple primary information receivers (PIRs) by passive reflection, but also simultaneously delivers additional information to a secondary information receiver (SIR) by embedding its information into the primary signals at the symbol level.
Two typical optimization problems, i.e., power minimization and quality-of-service (QoS) balancing, are investigated for the proposed IRS-based passive information transmission systems.
Simulation results demonstrate the feasibility of IRS-based passive information transmission and the effectiveness of our proposed algorithms, as compared to other benchmark schemes.
\end{abstract}

\begin{IEEEkeywords}
Intelligent reflecting surface (IRS), symbol-level precoding, passive information transmission, passive beamforming.
\end{IEEEkeywords}

\maketitle

\section{Introduction}
\vspace{0.2 cm}

During the past decade, various techniques have been developed to accommodate the rapidly increasing demands for high data rates and diverse quality-of-service (QoS).
Massive multi-input multi-output (MIMO), millimeter wave (mmWave) communication, and ultra-dense networks are three representative approaches for enhancing wireless network performance \cite{Zhang 2017}.
However, the required high hardware cost as well as the resulting increased energy consumption remain as roadblocks in their practical implementation.
Being a energy/spectrum/hardware-efficient solution for future wireless networks, intelligent reflecting surfaces (IRS) have attracted abundant attentions owing to its ability to tailor the radio environment in a cost-effective and hardware-efficient fashion \cite{Liaskos CM 2018}-\cite{Wu 2020}.

An IRS is a man-made two-dimensional (2D) surface composed of a large number of reconfigurable passive elements, \textcolor{black}{and is also referred to as dynamic metasurface antennas \cite{DMA}}.
Each element can independently manipulate the phase of the incident signals in a real-time programmable manner, thus collaboratively enabling adaptive reflection beamforming in three-dimensional (3D) space.
By intelligently controlling the signal reflection, an IRS can create a more favorable propagation environment, which used to be more or less out of the control of the system designer.
Furthermore, the reflection beamforming offers additional degrees of freedom (DoFs) for addressing severe channel fading and refining channel statistics.
Therefore, IRS is attracting steadily growing interest in both academia and industry.
In the past several years, researchers have devoted substantial efforts to exploring the potentials of IRS as a passive relay/reflector to greatly expand coverage, improve transmission quality, and assure security, \cite{Wu TWC 2019}-\cite{Yu JSAC 2020}, etc.
By judiciously designing the IRS phase-shifts, the signals reflected by the IRS and from other paths can coherently add up at intended receivers and/or cancel out at unintended receivers to improve system performance.
\textcolor{black}{Various advanced optimization algorithms and deep learning-based methods \cite{Huang JSAC 2020}, \cite{Yang TWC 2021} have been proposed for the designs of IRS-assisted systems.
In addition, IRS has also found some novel applications in holographic MIMO \cite{Huang WC 2020}, \cite{Wan TCOM 2021}, mobile edge computing, sensing and localization, etc. \cite{Renzo 19}, \cite{Wu 2020}.
}

In most of the applications mentioned above, the IRS is deployed as a \textit{passive reflector} to enhance performance by adaptively reflecting the incident signals, which are already modulated/precoded by an active transmitter.
Meanwhile, the novel concept of utilizing IRS as \textit{passive transmitters} was presented in \cite{Basar Access 2019}, where the IRS changes the parameters of the reflecting elements to modulate and transmit information symbols by exploiting an unmodulated carrier signal generated by a nearby radio-frequency (RF) signal generator.
The testbed platforms of utilizing IRS to realize a quadrature phase-shift-keying (QPSK) transmitter \cite{Tang CC 2019} and 8-PSK transmitter \cite{Tang EL 2019} validated this idea.
Specially, the IRS-based passive transmitter can realize virtual MIMO communication with only one RF chain and very cost-effective reflecting elements at the transmitter side, which thus makes it very promising for practical wireless networks due to the significantly reduced hardware complexity and energy consumption.
While the IRS-based single-stream transmitter was investigated in \cite{Basar Access 2019}, \cite{Tang CC 2019}, \cite{Tang EL 2019}, exploiting IRS to simultaneously transmit multiple data streams and serve multiple users has not been studied yet, which thus motivates this work.

In addition, there has been growing interest in combining the passive reflection and passive transmission capabilities of IRS.
In these joint applications, besides enhancing the quality of the primary signals using the IRS, one can also modulate and embed the secondary information into the primary signals by appropriately varying the IRS reflection coefficients.
\textcolor{black}{Such an IRS-based symbiotic radio system achieves higher spectrum efficiency by sharing the same spectrum resources with the primary and secondary information transmissions.
Compared with typical ambient backscatter communication based symbiotic radio systems \cite{Huynh CST 2018}, \cite{Liang TCCN 2020}, where the backscatter device is usually single-antenna, the abundant reflecting elements of IRS provide more DoFs to combat the double fading effect and enhance the quality of information transmissions.}
Specifically, in \cite{Yan WCL 2020}, \cite{Yan 2019}, the authors presented a joint passive beamforming and information transfer system, in which the secondary information is modulated by the on/off states of the IRS reflecting elements.
In \cite{Guo 2019}, the authors proposed a reflecting modulation scheme for an IRS-based passive transmitter.
In \cite{Zhang 2020}, the IRS operates as an Internet-of-Things device to transfer secondary information by jointly designing the active beamforming and passive reflecting.
While these works \cite{Basar Access 2019}, \cite{Tang CC 2019}, \cite{Tang EL 2019}, \cite{Yan WCL 2020}-\cite{Ma CL 20} explored the feasibility of using an IRS as a passive transmitter, the selection of reflection patterns are limited for the secondary information transmission, e.g., only two antipodal reflection patterns were considered in \cite{Zhang 2020}.
Therefore, the full potential of IRS has not been exploited in these works.
More importantly, in these existing designs, both primary and secondary receivers need to jointly detect the  primary and secondary information symbols, which causes high computational complexity to the receivers.
Furthermore, the more complicated case with passive information transmission to multiple receivers has not been investigated yet, which motivates us to develop this work.

We note that in IRS-based passive transmission schemes, each information symbol is modulated by varying the reflecting elements of the IRS, which is similar to the mechanism employed in symbol-level precoding \cite{CM ITSP 2015}-\cite{Liu ISPL 2017}.
In symbol-level precoding, the multi-antenna transmitter varies its precoder in a symbol-by-symbol fashion to turn the harmful multi-user interference (MUI) into constructive and beneficial signals.
Such methods can exploit both the spatial and symbol-level DoFs to significantly improve the symbol error-rate (SER) performance of multiuser systems.
Moreover, the receivers at the symbol-level precoding system can demodulate information by simple hard-decision since the optimizations at the transmitter side consider specific modulation type known at the receiver side.
\textcolor{black}{In addition, with current semiconductor technologies, fast positive-intrinsic-negative (PIN) or Schottky diodes can realize switching within nanoseconds \cite{Sedaghat CM 2016}. Existing testbed platforms \cite{Tang CC 2019}, \cite{Tang EL 2019} also validated the feasibility of adjusting IRS at the symbol-level speed.}
Inspired by these findings, we attempt to realize IRS-based passive information transmissions in this paper by exploiting symbol-level precoding technology, which provides symbiotic benefits from various aspects.
The main contributions in this paper are summarized as follows:
\textcolor{black}{\begin{itemize}
  \item Different from our previous work \cite{Liu TWC 20} on joint symbol-level precoding and IRS passive reflection design, in this paper, we use the idea behind symbol-level precoding to implement  IRS-based passive information transmissions in downlink multi-user multi-input single-output (MU-MISO) systems.
      Such a symbol-level precoding approach not only fully exploits the DoFs of massive reflecting elements to enhance the quality of information transmissions, but also allows the receivers to employ a very simple symbol detector, thus renders itself particularly appealing in practical implementation.
  \item For the case that the IRS works as a standalone passive transmitter to deliver information to multiple single-antenna users, we design the IRS phase-shifts to minimize the transmit power subject to a given set of QoS requirements.
      Efficient algorithms based on the Riemannian manifold optimization and the branch-and-bound algorithm are proposed to obtain continuous/high-resolution phase-shifts and low-resolution quantized phase-shifts, respectively.
      We also investigate the QoS balancing problem for a given transmit power budget.
  \item For the case that the IRS works as a joint passive reflector and transmitter, also known as symbiotic radio system, the IRS enhances the primary information transmissions from the multi-antenna base station (BS) to multiple single-antenna users, and simultaneously delivers secondary information to one additional user by embedding the secondary information into the primary signals.
      The power minimization and QoS balancing problems are also investigated by iteratively solving for the precoders and reflectors using efficient gradient projection-based and conjugate gradient-based algorithms.
  \item Finally, we provide extensive simulation results to demonstrate the feasibility of exploiting symbol-level precoding for passive information transmissions in IRS-based MU-MISO systems, and to illustrate the effectiveness of our proposed algorithms.
\end{itemize}}


\textit{Notation}: Boldface lower-case and upper-case letters indicate column vectors and matrices, respectively.
$(\cdot)^T$ and $(\cdot)^H$ denote the transpose and the transpose-conjugate operations, respectively.
$\mathbb{C}$ denotes the set of complex numbers.
$| a |$ and $\| \mathbf{a} \|$ are the magnitude of a scalar $a$ and the norm of a vector $\mathbf{a}$, respectively.
$y=f\langle x\rangle$ denotes that $y$ is a function of $x$.
$\angle{a}$ is the angle of complex-valued $a$.
$\mathfrak{R}\{\cdot\}$ and $\mathfrak{I}\{\cdot\}$ denote the real and imaginary part of a complex number, respectively.
$\text{diag}\{\mathbf{a}\}$ indicates a diagonal matrix whose diagonal terms are the elements of $\mathbf{a}$.
$\mathbf{A}(i,j)$ denotes the element of the $i$-th row and the $j$-th column of matrix $\mathbf{A}$, and  $\mathbf{a}(i)$ denotes the $i$-th element of vector $\mathbf{a}$.

\section{Passive Information Transmission System}
\vspace{0.2 cm}

\subsection{System Model}

\begin{figure}[!t]
\centering
\includegraphics[height = 1.95 in]{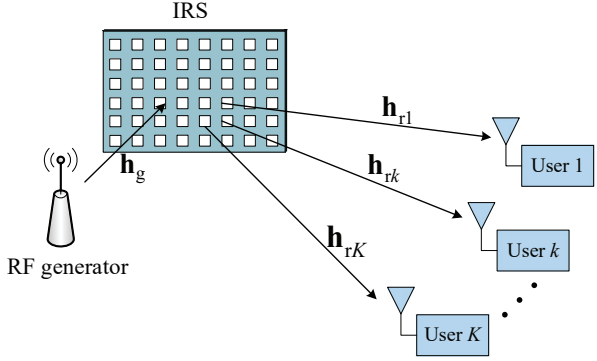}\vspace{0.1 cm}
\caption{An IRS-based MU-MISO passive information transmission system.}
\label{fig:system model}\vspace{0.1 cm}
\end{figure}

Consider an MU-MISO system as shown in Fig. \ref{fig:system model}, where the IRS equipped with $N$ reflecting elements simultaneously serves $K$ single-antenna users.
\textcolor{black}{Specifically, using the idea of symbol-level precoding, the IRS modulates the information symbols onto the high-frequency carrier signals, which are generated from a nearby RF signal generator, via correspondingly adjusting its reflection coefficients.
Since the information modulation and precoding are implemented at the IRS side, which only has passive components, we name this communication system as the passive information transmission system.}
As discussed in \cite{Basar Access 2019} and \cite{Basar ITC 2020}, the RF generator is sufficiently close to the IRS and utilizes a horn antenna to focus the signals on the IRS.
Therefore, the RF generator and IRS can be seen as a transmitter with only one RF chain and multiple reflecting elements, which realizes a virtual MIMO communication with much reduced hardware complexity and cost.
Denote $\bm{\theta}\in \mathbb{C}^N$ as the vector containing the IRS reflection coefficients.
The received baseband signal at the $k$-th user can be written as\footnote{The direct link between the RF generator and the users is ignored since the pure RF signal does not contain information (i.e., a constant baseband value) and can be easily removed at the users. Besides, the RF generator mainly beams the energy towards the IRS rather than the randomly distributed users.}
\be
r_k = \sqrt{P}\mathbf{h}_{\text{r}k}^H\bm{\Theta}\mathbf{h}_\text{g} + n_k,
\ee
where $P$ is the transmit power of the RF generator, $\bm{\Theta} \triangleq \text{diag}\{\bm{\theta}\}$, $\mathbf{h}_{\text{r}k} \in \mathbb{C}^N$ is the channel vector from the IRS to the $k$-th user, $\mathbf{h}_{\text{g}}$ is the channel vector from the RF generator to the IRS, and $n_k \sim \mathcal{CN}(0,\sigma^2)$ is additive white Gaussian noise (AWGN) at the $k$-th user.
Given the availability of various channel estimation approaches \cite{Wu 2020}, \cite{You 2019}-\cite{Wei TCOM 2021}, perfect channel state information (CSI) is assumed to be available at the IRS.
For simplicity, we denote the equivalent channel from the IRS to the $k$-th user as
$\mathbf{h}_k^H \triangleq \mathbf{h}_{\text{r}k}^H\text{diag}\{\mathbf{h}_{\text{g}}\}$.
Then, the received signal for the $k$-th user can be rewritten as
\be
r_k = \sqrt{P}\mathbf{h}_k^H\bm{\theta} + n_k.
\ee

To realize passive information transmissions, the IRS modulates the symbols by changing the reflection vector $\bm{\theta}$ according to the symbols to be transmitted \cite{Basar Access 2019}.
We assume that the desired symbols for all users are independently $\Omega$-PSK modulated\footnote{It is noted that symbol-level precoding is related to the modulation type. The designs throughout this paper focus on PSK modulations. The designs for quadrature amplitude modulation (QAM) are left for our future work.}.
As a result, there are $\Omega^K$ possible combinations of symbols that could be transmitted to the $K$ users.
We let $\mathbf{s}_m \triangleq [s_{m,1},\ldots,s_{m,K}]^T$ for $m = 1,\ldots,\Omega^K$ represent all of the possible symbol vectors.
To transmit $\mathbf{s}_m$, an appropriate phase vector $
\bm{\theta}_m$ must be designed, so that the received signal at the $k$-th user
\be
r_{m,k} = \sqrt{P}\mathbf{h}_k^H\bm{\theta}_m + n_k
\ee
can be properly decoded according to the pre-known constellation information.
In the sequel, we will design the non-linear mapping from $\mathbf{s}_m$ to $\bm{\theta}_m$ using the  constructive symbol-level precoding \cite{CM ITSP 2015}-\cite{Liu ISPL 2017}.
Considering the ideal reflection model with unit-modulus and continuous phase shifts, each reflecting element should satisfy $|\bm{\theta}_m(n)| = 1, \forall m, n$.

\begin{figure}[!t]
\centering
\subfigure[]{
\begin{minipage}{4.2 cm}
\centering
\includegraphics[height = 5.4 cm]{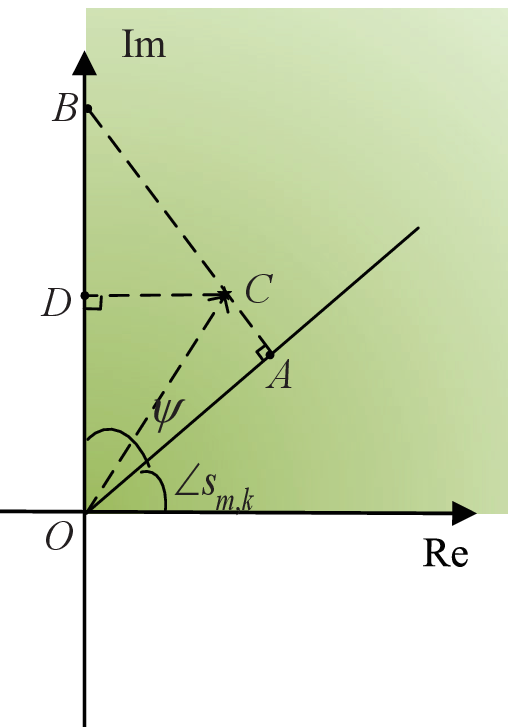}\vspace{0.1 cm}
\label{fig:CR1}
\end{minipage}}
\subfigure[]{
\begin{minipage}{4.2 cm}
\centering
\includegraphics[height = 5.4 cm]{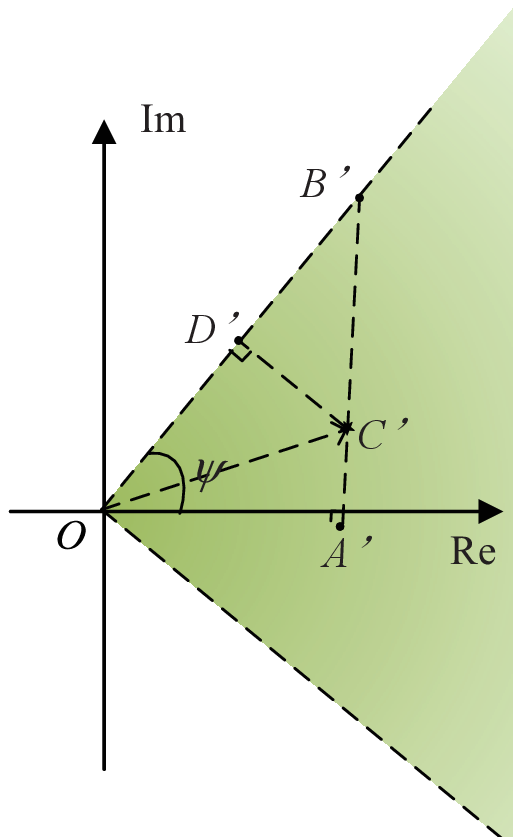}\vspace{0.1 cm}
\label{fig:CR2}
\end{minipage}}\vspace{0.1 cm}
\caption{Symbol-level precoding design for QPSK signals.}
\label{fig:CR}
\end{figure}

Using knowledge of the symbol vector $\mathbf{s}_m$ to be transmitted, constructive interference (CI) symbol-level precoding converts the MUI into constructive components that push the received noise-free signals away from their corresponding decision boundaries, which greatly improves the detection performance.
Thus, the Euclidean distance between the received noise-free signal and its corresponding closest decision boundary is adopted as the metric to measure the users' QoS.
\textcolor{black}{More detailed and comprehensive descriptions about constructive interference and symbol-level precoding can be found in \cite{CM ITSP 2015}-\cite{Liu ISPL 2017}.}
In order to explicitly demonstrate this metric, we take QPSK modulated symbols as an example, as shown in Fig. \ref{fig:CR}.
Specifically, we assume $s_{m,k} = e^{\jmath\pi/4}$ is the desired symbol of the $k$-th user, the positive halves of the $x$ and $y$ axes are the decision boundaries for $s_{m,k}$, the shaded green area is its decision region (i.e., CI region), $\overrightarrow{OC}=\sqrt{P}\mathbf{h}_k^H\bm{\theta}_m$ is the received noise-free signal, $\psi=\pi/\Omega = \pi/4$ is the half angle of the decision region, and $|\overrightarrow{CD}|$ is the Euclidean distance between point $C$ and the corresponding nearest decision boundary.
We see that $s_{m,k}$ can be correctly detected at the $k$-th user when the received signal $r_{m,k}$ lies in the green region.
In order to improve the robustness to the noise, the phase vector $\bm{\theta}_m$ should be designed to render the received noise-free signal $\overrightarrow{OC}$ as farther away from its decision boundaries as possible.
Therefore, $|\overrightarrow{CD}|$ is usually used as the metric to evaluate the QoS of communication since it determines the SER performance.
To derive the expression of $|\overrightarrow{CD}|$, we rotate the diagram clockwise by $\angle s_{m,k}$ degrees as shown in Fig. \ref{fig:CR2}, where \be\overrightarrow{OC'} = \widetilde{r}_{m,k} = \sqrt{P} \mathbf{h}_k^H\bm{\theta}_me^{-\jmath\angle s_{m,k}}.\ee
Then, the Euclidean distance to evaluate the QoS of communication is given by
\begin{equation}
\begin{aligned}
|C'D'| &= |C'B'|\cos\psi = (|A'B'|-|A'C'|)\cos\psi \\
& = \left(\mathfrak{R}\{\overrightarrow{OC'}\}\tan\psi -
|\mathfrak{I}\{\overrightarrow{OC'}\}|\right)\cos\psi \\
\label{eq:distance}
& = \mathfrak{R}\left\{\widetilde{r}_{m,k}\right\}\sin\psi-\left|\mathfrak{I}
\left\{\widetilde{r}_{m,k}\right\}\right|\cos\psi.
\end{aligned}
\end{equation}
The QoS constraint can thus be expressed as
\be
\mathfrak{R}\left\{\widetilde{r}_{m,k}\right\}\sin\psi-\left|\mathfrak{I}
\left\{\widetilde{r}_{m,k}\right\}\right|\cos\psi \geq \alpha_k, \forall m, k,
\ee
where $\alpha_k$ is the preset QoS requirement for the $k$-th user.

\subsection{Design for Power Minimization}

In this subsection, we investigate the power minimization problem, which aims to minimize transmit power at the RF generator while satisfying the QoS requirements of all users.
Accordingly, the optimization problem is formulated as
\begin{subequations}
\label{eq:PIT PM}
\begin{align}
\label{eq:PIT PM obj}
&\underset{\bm{\theta}_m,\forall m, P}{\min}~~P \\
\label{eq:PIT PM c1}
&~\text{s.t.}~~~ \mathfrak{R}\left\{\widetilde{r}_{m,k}\right\}\sin\psi-\left|\mathfrak{I}
\left\{\widetilde{r}_{m,k}\right\}\right|\cos\psi \geq \alpha_k, \forall m,k,\\
&~~~~~~~~\widetilde{r}_{m,k} = \sqrt{P} \mathbf{h}_k^H\bm{\theta}_me^{-\jmath\angle s_{m,k}},
\forall m,k, \\
&~~~~~~~\left|\bm{\theta}_m(n)\right| = 1, \forall m,n.
\end{align}
\end{subequations}
It can be observed that different from the constant envelope symbol-level precoding approach in \cite{Amadori TWC 2017}, we utilize the IRS to realize passive beamforming and simultaneously consider all possible precoders in one channel coherence time.
To efficiently solve this non-convex multivariate optimization problem, we propose to convert it into a univariate problem, which facilitates us to develop an efficient algorithm.

First, after dividing both sides of the QoS constraint (\ref{eq:PIT PM c1}) by $\alpha_k\sqrt{P}$, we have
\be\label{eq:for p}
\frac{1}{\sqrt{P}} \leq \frac{1}{\alpha_k}\left[\mathfrak{R}\left\{\widehat{r}_{m,k}\right\}\sin\psi-\left|\mathfrak{I}
\left\{\widehat{r}_{m,k}\right\}\right|\cos\psi\right], \forall m, k,
\ee
where $\widehat{r}_{m,k} \triangleq \mathbf{h}_k^H\bm{\theta}_me^{-\jmath\angle s_{m,k}}$.
By introducing an auxiliary variable $t \triangleq \frac{1}{\sqrt{P}}$, the power minimization problem (\ref{eq:PIT PM}) can be converted to
\begin{subequations}
\label{eq:PIT PM QoS t}
\begin{align}
\label{eq:PIT PM QoS obj t}
&\underset{\bm{\theta}_m,\forall m,t}{\max}~~t \\
&~~\text{s.t.}~~~~~t \leq \frac{1}{\alpha_k}\Big[\mathfrak{R}\left\{\widehat{r}_{m,k}\right\}\sin\psi-\left|\mathfrak{I}
\left\{\widehat{r}_{m,k}\right\}\right|\cos\psi\Big], \forall m,k,\\
&\hspace{1.2 cm}~ \widehat{r}_{m,k} = \mathbf{h}_k^H\bm{\theta}_me^{-\jmath\angle s_{m,k}}, \forall m, k,\\
&\hspace{1.1 cm}~ \left|\bm{\theta}_m(n)\right| = 1, \forall m,n.
\end{align}
\end{subequations}
The optimization problem (\ref{eq:PIT PM QoS t}) can be further equivalently rewritten as a max-min problem as
\begin{subequations}
\label{eq:PIT PM QoS}
\begin{align}
\label{eq:PIT PM QoS obj}
\underset{\bm{\theta}_m,\forall m}{\max}~~&\underset{m,k}{\min}~~
\frac{1}{\alpha_k}\Big[\mathfrak{R}\left\{\widehat{r}_{m,k}\right\}\sin\psi-\left|\mathfrak{I}
\left\{\widehat{r}_{m,k}\right\}\right|\cos\psi\Big] \\
&~\text{s.t.}~~\widehat{r}_{m,k} = \mathbf{h}_k^H\bm{\theta}_me^{-\jmath\angle s_{m,k}}, \forall m,k,\\
\label{eq:PIT PM QoS c2}
&~~~~~~\left|\bm{\theta}_m(n)\right| = 1, \forall m,n.
\end{align}
\end{subequations}
Obviously, problem (\ref{eq:PIT PM QoS}) is also difficult to solve, not only because of the large number of variables, but also the non-differentiable absolute and minimum value functions in the objective (\ref{eq:PIT PM QoS obj}), and the non-convex unit-modulus constraint (\ref{eq:PIT PM QoS c2}) of the IRS.
In the following, we propose to divide the large-scale optimization problem into several  sub-problems.
For each sub-problem, an approximate differentiable objective is derived, and the non-convex constraint is tackled using a manifold-based algorithm.

Since the optimization of (\ref{eq:PIT PM QoS}) for different $\bm{\theta}_m, m = 1, \ldots, \Omega^K$, is independent with respect to $m$, we can equivalently divide this large-scale problem into $\Omega^K$ sub-problems.
The $m$-th sub-problem is rewritten as
 \begin{subequations}
\label{eq:PIT PM QoS m}
\begin{align}
\label{eq:PIT PM QoS m obj}
\underset{\bm{\theta}_m}{\min}~~&\underset{i}{\max}~~
\left|\mathfrak{I}\left\{\mathbf{a}_i^H\bm{\theta}_m\right\}\right|\cos\psi-
\mathfrak{R}\left\{\mathbf{a}_i^H\bm{\theta}_m\right\}\sin\psi \\
\label{eq:PIT PM QoS m c}
&~\text{s.t.}~~~\left|\bm{\theta}_m(n)\right| = 1, \forall n,
\end{align}
\end{subequations}
where $\mathbf{a}_i^H \triangleq \frac{1}{\alpha_k}\mathbf{h}_k^He^{-\jmath\angle s_{m,k}}, i = K(m-1)+k$.
Then, in order to remove the absolute value function and form a more concise objective function, we utilize the property that $|x| = \max\{x,-x\}$ for a scalar $x$ together with some basic linear algebra laws to re-formulate the objective (\ref{eq:PIT PM QoS m obj}) as
\be\label{eq:max fi gi}
\left|\mathfrak{I}\left\{\mathbf{a}_i^H\bm{\theta}_m\right\}\right|\cos\psi-
\mathfrak{R}\left\{\mathbf{a}_i^H\bm{\theta}_m\right\}\sin\psi = \max\{f_i, g_i\},
\ee
where $f_i$ and $g_i$ are defined as
\begin{subequations}\label{eq:fg}
\begin{align}
\label{eq:fi}
f_i &\triangleq \mathfrak{I}\left\{\mathbf{a}_i^H\bm{\theta}_m\right\}\cos\psi-
\mathfrak{R}\left\{\mathbf{a}_i^H\bm{\theta}_m\right\}\sin\psi \non\\
&= \mathfrak{R}\left\{\mathbf{b}_i^H\bm{\theta}_m\right\}, \forall i,\\
\label{eq:gi}
g_i &\triangleq -\mathfrak{I}\left\{\mathbf{a}_i^H\bm{\theta}_m\right\}\cos\psi-
\mathfrak{R}\left\{\mathbf{a}_i^H\bm{\theta}_m\right\}\sin\psi \non \\
&= \mathfrak{R}\left\{\mathbf{c}_i^H\bm{\theta}_m\right\}, \forall i.
\end{align}
\end{subequations}
For notational conciseness in the following algorithm development, in (\ref{eq:fi}) and (\ref{eq:gi}), we have defined
\begin{subequations}\label{eq:bcuv}
\begin{align}
\mathbf{b}_i^H &\triangleq -\mathbf{a}_i^H\sin\psi + \mathbf{a}_i^H e^{-\jmath\frac{\pi}{2}}\cos\psi,\\
\mathbf{c}_i^H &\triangleq -\mathbf{a}_i^H\sin\psi - \mathbf{a}_i^H e^{-\jmath\frac{\pi}{2}}\cos\psi.
\end{align}
\end{subequations}
Then, exploiting the well-known \textit{log-sum-exp} approximation, the objective (\ref{eq:PIT PM QoS m obj}) is converted to
\be\begin{aligned}\label{eq:PIT PM h}
\underset{i}{\max}~\{f_i, g_i\}\lessapprox \varepsilon \log \sum_{i=K(m-1)+1}^{mK}
\big[\exp(f_i/\varepsilon)+\exp(g_i/\varepsilon)\big],
\end{aligned}\ee
where $\varepsilon > 0$ is a relatively small number to maintain the approximation.

After obtaining the smooth and differentiable objective (\ref{eq:PIT PM h}) for problem (\ref{eq:PIT PM QoS m}), the unit-modulus constraint (\ref{eq:PIT PM QoS m c}) of IRS becomes the major challenge.
The non-convex relaxation and alternating optimization based algorithms are very popular to solve this problem.
However, the relaxation-based algorithms may suffer significant performance loss and the alternating optimization algorithms may have a slow convergence.
To avoid these issues, we adopt the Riemannian-manifold-based algorithm, which directly solves this problem on the original feasible space instead of a relaxed convex version, and thus is able to provide a locally optimal solution with fast convergence \cite{CG}.
Constraint (\ref{eq:PIT PM QoS m c}) forms an $N$-dimensional complex circle manifold
\begin{equation}
\label{eq:search space}
\mathcal{M}_\text{cc} = \left\{\bm{\theta}_m \in \mathbb{C}^{N}:\bm{\theta}^*_m(n)\bm{\theta}_m(n) = 1, \forall n\right\},
\end{equation}
which is a smooth Riemannian manifold equipped with an inner product defined on the tangent space:
\begin{equation}\label{eq:tangent space}
T_{{\bm{\theta}}_m}\mathcal{M}_\text{cc} = \left\{\mathbf{p} \in \mathbb{C}^{N}: \mathfrak{R}\left\{\mathbf{p}\odot\bm{\theta}_m^*\right\} = \mathbf{0}_N, \forall n\right\}.
\end{equation}
Then, problem (\ref{eq:PIT PM QoS m}) can be rewritten as
\begin{equation}
\label{eq:hm}
\underset{\bm{\theta}_m\in\mathcal{M}_\text{cc}}{\min}~h\langle\bm{\theta}_m\rangle = \varepsilon \log \sum_{i=K(m-1)+1}^{mK}
\big[\exp(f_i/\varepsilon)+\exp(g_i/\varepsilon)\big]
\end{equation}
\nid which is an unconstrained optimization problem on the Riemannian space $\mathcal{M}_\text{cc}$.
Since each point on the manifold has a neighborhood homeomorphic to Euclidean space, the gradients of cost functions, distances, angles, etc., have counterparts on the Riemannian space, and efficient algorithms developed on the Euclidean space can be readily extended to the manifold space, e.g., the conjugate gradient (CG) algorithm.
Therefore, in the following we use the CG algorithm on the Riemannian space, referred to as the Riemannian conjugate gradient (RCG) algorithm, to solve this problem.

To facilitate the RCG algorithm, we first derive the Euclidean gradient of $h\langle{\bm{\theta}}_m\rangle$ by
\be
\nabla\;h\langle\bm{\theta}_m\rangle =
\frac{\sum_{i=K(m-1)+1}^{mK}
\big[\exp(f_i/\varepsilon)\mathbf{b}_i+\exp(g_i/\varepsilon)\mathbf{c}_i\big]}{\sum_{i=K(m-1)+1}^{mK}
\big[\exp(f_i/\varepsilon)+\exp(g_i/\varepsilon)\big]}.
\ee
Then, the Riemannian gradient $\text{grad}\;h\langle\bm{\theta}_m\rangle$ can be obtained by projecting the Euclidean gradient $\nabla\;h\langle\bm{\theta}_m\rangle$ onto its corresponding Riemannian tangent space as:
\be\label{eq:grad}
\begin{aligned}
\text{grad}\;h\langle\bm{\theta}_m\rangle &= \text{Proj}_{\bm{\theta}_m}\nabla h\langle\bm{\theta}_m\rangle \\
& = \nabla h\langle\bm{\theta}_m\rangle - \mathfrak{R}\left\{\nabla h\langle\bm{\theta}_m\rangle\odot\bm{\theta}_m^*\right\}\odot\bm{\theta}_m.
\end{aligned}
\ee
Thus, in the $p$-th iteration of the RCG algorithm, the search direction $\mathbf{d}_p$ is
\begin{equation}
\label{eq:dp}
\mathbf{d}_p = -\text{grad}\;h\langle\bm{\theta}_{m,p-1}\rangle + \eta_p \mathbf{d}_{p-1}^\text{t},
\end{equation}
where $\bm{\theta}_{m,p-1}$ is the solution in the $(p-1)$-th iteration, and $\eta_p$ is the Polak-Ribiere parameter \cite{CG}.
Since the Riemannian gradient and search direction for the $(p-1)$-th iteration lie in different tangent spaces, an additional Riemannian transport operation is needed to map $\mathbf{d}_{p-1}$ into the tangent space of $\text{grad}\;h\langle\bm{\theta}_{m,p-1}\rangle$ denoted by $\mathbf{d}_{p-1}^\text{t}$.
After choosing the step size $\xi_p$ using the Armijo backtracking line search method \cite{CG}, the $p$-th update is given by
\begin{equation}
\label{eq:retr}
\bm{\theta}_{m,p+1} = \text{Retr}_{\bm{\theta}_m}\left(\bm{\theta}_{m,p}+\xi_p\mathbf{d}_p\right),
\end{equation}
where $\text{Retr}_{\bm{\theta}_m}\left(\cdot\right)$ indicates the retraction operation, mapping the point on the tangent space to the manifold.

Based on the above analysis, the locally optimal solution to each $\bm{\theta}_m^\star$ can be obtained using the RCG algorithm as summarized in Algorithm 1.
Then, the minimum required power can be calculated by substituting $\bm{\theta}_m^\star, \forall m$ to (\ref{eq:for p}).
\textcolor{black}{Besides, as proved in Theorem 4.3.1 \cite{CG}, Algorithm 1 is guaranteed to converge to a critical point where the Riemannian gradient is equal to zero.}

\begin{algorithm}[!t]
\begin{small}
\caption{RCG algorithm to obtain $\bm{\theta}_m^\star$}
\label{alg:RCG}
\begin{algorithmic}[1]
\REQUIRE $h\langle{\bm{\theta}}_m\rangle$.
\ENSURE $\bm{\theta}_m^\star$.
    \STATE {Initialize ${\bm{\theta}}_{m,0} \in \mathcal{M}_\text{cc}$, $\mathbf{d}_0 = -\text{grad}_{{\bm{\theta}}_m}h\langle{\bm{\theta}}_{m,0}\rangle$.}
    \STATE{\textbf{Repeat}}
        \STATE{\hspace{0.4 cm}Choose Polak-Ribiere parameter $\eta_{p}$ \cite{CG}.}
        \STATE{\hspace{0.4 cm}Calculate search direction $\mathbf{d}_{p}$ by (\ref{eq:dp}).}
        \STATE{\hspace{0.4 cm}Calculate step size $\xi_p$ \cite{CG}.}
        \STATE{\hspace{0.4 cm}Obtain the update ${\bm{\theta}}_{m,p+1}$ by (\ref{eq:retr}).}
        \STATE{\hspace{0.4 cm}Calculate gradient $\text{grad}_{{\bm{\theta}}_m}h\langle{\bm{\theta}}_{m,p}\rangle$ by (\ref{eq:grad}).} \STATE{\textbf{Until convergence}}
    \end{algorithmic}\end{small}
\end{algorithm}

\subsection{Design for QoS Balancing}

In this subsection, we consider the QoS balancing problem, which aims to maximize the minimum weighted QoS performance for a given transmit power.
The optimization problem is formulated as
\begin{subequations}
\label{eq:PIT QoS}
\begin{align}
\label{eq:PIT QoS obj}
\underset{\bm{\theta}_m,\forall m}{\max}~~&\underset{m,k}{\min}~~
\rho_k\Big[\mathfrak{R}\left\{\widetilde{r}_{m,k}\right\}\sin\psi-\left|\mathfrak{I}
\left\{\widetilde{r}_{m,k}\right\}\right|\cos\psi\Big] \\
&~\text{s.t.}~~\widetilde{r}_{m,k} = \sqrt{P} \mathbf{h}_k^H\bm{\theta}_me^{-\jmath\angle s_{m,k}}, \forall m,k,\\
&~~~~~~\left|\bm{\theta}_m(n)\right| = 1, \forall m,n,
\end{align}
\end{subequations}
where $P$ is the maximum transmit power and $\rho_k > 0$ is the weight coefficient for the $k$-th user.
It can be seen that this problem is very similar to problem (\ref{eq:PIT PM QoS}).
Therefore, this QoS balancing problem can be solved with the algorithm proposed in Sec. II-B by setting  $\alpha_k = \frac{1}{\rho_k\sqrt{P}}, \forall k$ in (\ref{eq:PIT PM QoS}).

\subsection{Design for Low-Resolution IRS}

Since an IRS with infinite/high-resolution phase shifters would inevitably require higher hardware complexity and cost, low-resolution phase shifters are practically appealing.
Thus, in this subsection, we investigate solutions for the case of low-resolution IRS.

With the continuous solution $\bm{\theta}_m^\star$ obtained as in Sec. II-B, a direct quantization operation as in \cite{Wu 2020 TCOM} can be easily applied to seek the nearest discrete phase value by
\begin{equation}
\label{eq: theta mn}
\angle{\bm{\theta}_m^B(n)} = \left[\frac{\angle{\bm{\theta}_m^\star(n)}}{\Delta}\right] \times \Delta,
\end{equation}
where $\Delta \triangleq \frac{2\pi}{2^B}$ is the resolution of each reflecting elements controlled by $B$ bits, and $\left[\cdot\right]$ indicates the rounding operation.
However, this method provides a suboptimal solution due to the quantization error, which may cause severe performance degradation for very low-resolution cases, e.g., 1-bit and 2-bit cases \cite{Wu 2020}.

Thus, we investigate obtaining the optimal solutions by converting the optimization problem into a mixed-integer nonlinear program (MINLP) and solving it with an off-the-shelf algorithm.
The low-resolution phase-shifts can be rewritten as
\begin{equation}
\bm{\theta}_m^B = \mathbf{Q}_m \mathbf{q},
\end{equation}
where the auxiliary vector $\mathbf{q} \triangleq [e^{\jmath\Delta},e^{\jmath2\Delta},\ldots,e^{\jmath2\pi}]^T$ contains all the possible phase values, $\mathbf{Q}_m \in \left\{0,1\right\}^{N\times2^B}$ has only one non-zero element per row, and $\mathbf{Q}_m(n,j) = 1$ indicates the $n$-th element in $\bm{\theta}_m$ is $\mathbf{q}(j)$.
Then, the optimization problem for the low-resolution IRS case is re-formulated as
\begin{subequations}
\label{eq:p1 1bit}
\begin{align}
\underset{\mathbf{Q}_m}{\min}~&\underset{k}{\max}~
\left|\mathfrak{I}\left\{\mathbf{a}_i^H\mathbf{Q}_m\mathbf{q}\right\}\right|\cos\psi-
\mathfrak{R}\left\{\mathbf{a}_i^H\mathbf{Q}_m\mathbf{q}\right\}\sin\psi \\
~&\text{s.t.}~~~\sum_{j=1}^{2^B}\mathbf{Q}_m(n,j) = 1, \forall n,\\
&~~~~~~~\mathbf{Q}_m(n,j) \in \left\{0,1\right\}, \forall n, j,
\end{align}
\end{subequations}
which is an MINLP problem and can be efficiently solved using the well-known branch-and-bound algorithm \cite{Narendra ITC 1977}.
The details of this well-known algorithm are omitted for brevity.
When the optimal $\mathbf{Q}_m^\star$ for problem (\ref{eq:p1 1bit}) is found, the optimal low-resolution reflection coefficients $\bm{\theta}_m^{B\star}$ can be constructed as
\begin{equation}
\bm{\theta}_m^{B\star} = \mathbf{Q}_m^\star \mathbf{q}.
\end{equation}
However, considering the required high computational complexity, the branch-and-bound algorithm is only suitable for the 1-bit or 2-bit cases.

In order to provide better performance than direct quantization and lower complexity than the branch-and-bound algorithm, we further propose an efficient heuristic algorithm to successively seek the conditionally optimal low-resolution solutions.
Assuming that only the $n$-th element of $\bm{\theta}_m$ is unknown, the optimization problem is formulated as\begin{subequations}\label{eq:heuristic}\begin{align}
\underset{\bm{\theta}_m(n)}{\min}~&\underset{k}{\max}~~
\left|\mathfrak{I}\left\{c_i+a_{i,n}\bm{\theta}_m(n)\right\}\right|\cos\psi \non\\
&~~~~~~~~~~~~~-
\mathfrak{R}\left\{c_i+a_{i,n}\bm{\theta}_m(n)\right\}\sin\psi \\
~&\text{s.t.}~~~~ \bm{\theta}_m(n)\in \left\{e^{\jmath\Delta},e^{\jmath2\Delta},\ldots,e^{\jmath2\pi}\right\},
\end{align}
\end{subequations}
where $c_i \triangleq \sum_{j=1,j\neq n}^{N}a_{i,j}\bm{\theta}_m(j)$, and $a_{i,j}$ denotes the $j$-th element of $\mathbf{a}_i^H$.
Since in this case the number of feasible discrete values is not very large, an exhaustive search for the solution to (\ref{eq:heuristic}) is affordable.
With the obtained continuous $\bm{\theta}_m^\star$ as the initial value, we iteratively solve (\ref{eq:heuristic}) for each element of the IRS until convergence is achieved.

\subsection{Computational Complexity Analysis}

The worst-case complexity of the RCG algorithm to obtain $\bm{\theta}_m^\star$ is of order $\mathcal{O}\{N^{1.5}\}$ \cite{CG}.
Thus, the total complexity for the power minimization and QoS balancing problems with continuous IRS is of order $\mathcal{O}\{\Omega^KN^{1.5}\}$.
For low-resolution IRS, the complexity of the quantization operation in (\ref{eq: theta mn}) is of order $\mathcal{O}\{2N\}$, which can be neglected compared with the complexity required to obtain the continuous solution.
Thus, the total complexity for the low-resolution IRS scenario using quantization is also of order $\mathcal{O}\{\Omega^KN^{1.5}\}$.
For $B$-bit IRS using the branch-and-bound algorithm, the optimization problem is a $2^B$-dimensional integer program with $N$ variables, whose complexity is of order $\mathcal{O}\{2^{3.5B}N^{2.5}+2^{2.5B}N^{3.5}\}$ \cite{Narendra ITC 1977}.
Thus, the total complexity for this scenario is of order $\mathcal{O}\{\Omega^K(2^{3.5B}N^{2.5}+2^{2.5B}N^{3.5})\}$.
The computational complexity of the branch-and-bound algorithm is exponential in $B$, and is obviously much higher than direct quantization.
The complexity of the heuristic algorithm is of order $\mathcal{O}\{\Omega^K[N^{1.5}+K(2^B+N)]\}$, which is slightly higher than direct quantization but much lower than the branch-and-bound algorithm.

\section{Joint Passive Reflection and Information Transmission System}
\vspace{0.2 cm}

\subsection{System Model}

\begin{figure}[!t]
\centering
\includegraphics[width = 0.45\textwidth]{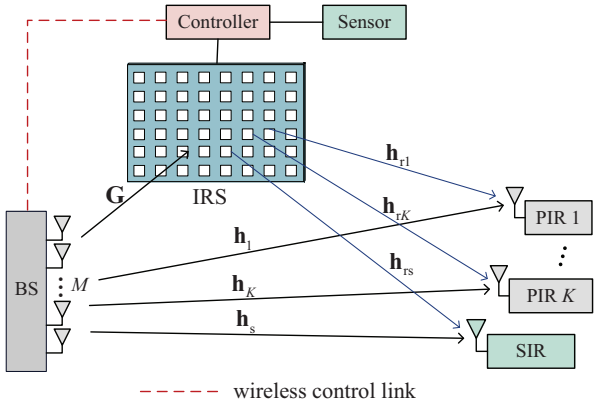}\vspace{0.1 cm}
\caption{Joint passive reflection and information transmission system, where IRS serves as a reflector and transmitter.}
\label{fig:system model2}
\vspace{0.1 cm}
\end{figure}

In this section, we introduce the joint passive reflection and information transmission system as shown in Fig. \ref{fig:system model2}, where the IRS is a dual-functional device, i.e., it combines the passive reflection function as in the traditional IRS-assisted downlink MU-MISO systems \cite{Wu TWC 2019} and the passive information transmission function as described in the previous section.
In particular, the IRS with $N$ passive reflecting elements enhances the primary information transmissions from an $M$-antenna BS to $K$ single-antenna primary information receivers (PIRs) by adjusting its reflection coefficients.
Meanwhile, it also transmits the secondary information to one secondary information receiver (SIR) by selecting the proper reflection.
The secondary information, e.g., temperature, light, or humidity, etc., is collected from the surroundings by a sensor or Internet of Things (IoT) device.
We assume that the $K$ PIRs and one SIR\footnote{\textcolor{black}{We should emphasize that the following derivations can be easily generalized to the scenario that there are multiple SIRs.}} are separated receivers.

We emphasize that in the considered joint passive reflection and information transmission system, the primary function of the IRS is to enhance the active information transmissions from the BS to PIRs by passive reflection.
The passive information transmission from the IRS to the SIR is the secondary task, which requires very low transmission rate.
To perform the above dual functions, IRS deployment and operation mechanism are quite different compared to the standalone passive information transmission system in the previous section.
In the following, we present detailed system models for both primary and secondary information transmissions.

For the primary information transmissions, the active precoding at the BS and passive reflection at the IRS are jointly optimized.
Specifically, the non-linear symbol-level precoding is employed since it not only enhances the primary information transmissions by converting harmful MUI into useful signals, but also provides additional symbol-level DoFs for embedding secondary information.
As before, assume that the primary information symbols for the $K$ PIRs are independent $\Omega$-PSK modulated symbols, and let $\mathbf{s}_m \triangleq [s_{m,1},\ldots,s_{m,K}]^T, m = 1, \ldots, \Omega^{K}$ represent all possible symbol combinations for the $K$ PIRs.
The precoder for transmitting $\mathbf{s}_m$ is denoted by $\mathbf{x}_m \in \mathbb{C}^M$ and the signal received by the $k$-th PIR is
\be
y_{m,k} = (\mathbf{h}_k^H+\mathbf{h}_{\text{r}k}^H\bm{\Theta}\mathbf{G})\mathbf{x}_m + n_k, m = 1, \ldots, \Omega^{K},
\ee
where $\mathbf{h}_k\in \mathbb{C}^{M}$, $\mathbf{h}_{\text{r}k}\in \mathbb{C}^{N}$, and $\mathbf{G}\in \mathbb{C}^{N\times M}$ are the channels from the BS to the $k$-th PIR, from the IRS to the $k$-th PIR, and from the BS to IRS, respectively, $\bm{\Theta} \triangleq \text{diag}\{\bm{\theta}\}$ is the reflection matrix, and $n_k \sim \mathcal{CN}(0,\sigma^2)$ is AWGN at the $k$-th PIR.
Similar to Sec. II, the PIRs detect their desired symbols by simple hard decisions according to the pre-known constellation information.
In addition, the QoS constraint for the PIRs is written as
\be\label{eq: PR QoS pre}
\mathfrak{R}\left\{\widetilde{y}_{m,k}\right\}\sin\psi-\left|\mathfrak{I}
\left\{\widetilde{y}_{m,k}\right\}\right|\cos\psi \geq \alpha_k, \forall m, k,
\ee
where $\widetilde{y}_{m,k} = (\mathbf{h}_k^H+\mathbf{h}_{\text{r}k}^H\bm{\Theta}\mathbf{G})
\mathbf{x}_{m}e^{-\jmath\angle s_{m,k}}$ is the rotated noise-free signal as before.

For the secondary passive information transmission, we propose to embed the secondary information into the primary signals using the extra DoFs available in the reflection coefficients of the IRS.
\textcolor{black}{Since only a small amount of data is transmitted by the sensor in general, the rate of the secondary transmission is usually much lower than the primary transmissions.
Therefore, we propose to embed ``binary'' secondary information symbols into the $L$ primary signals by selecting the reflection matrix as either $\bm{\Theta}_0\triangleq\text{diag}\{\bm{\theta}_0\}$ or $\bm{\Theta}_1\triangleq\text{diag}\{\bm{\theta}_1\}$, depending on whether a ``0'' or ``1'' is being sent by the sensor.}
In other words, the reflection matrix of the IRS is $\bm{\Theta}_0$ during $L$ time slots when the transmitted binary secondary symbol is ``0", and $\bm{\Theta}_1$ when it is ``1".
A smaller $L$ will provide a higher secondary symbol rate, but usually also results in worse SER performance.
Meanwhile, since the BS does not have access to the secondary information and the secondary information transmission should avoid to affect the primary information transmissions, both reflection matrices $\bm{\Theta}_0$ and $\bm{\Theta}_1$ should satisfy the QoS constraint for the PIRs in (\ref{eq: PR QoS pre}).
The QoS constraint for the PIRs is thus re-written as
\be\label{eq: PR QoS}
\mathfrak{R}\left\{\widetilde{y}_{m,k,0/1}\right\}\sin\psi-\left|\mathfrak{I}
\left\{\widetilde{y}_{m,k,0/1}\right\}\right|\cos\psi \geq \alpha_k, \forall m, k,
\ee
where $\widetilde{y}_{m,k,0/1} = (\mathbf{h}_k^H+\mathbf{h}_{\text{r}k}^H\bm{\Theta}_{0/1}\mathbf{G})
\mathbf{x}_{m}e^{-\jmath\angle s_{m,k}}$ and the subscript $0/1$ denotes whether the reflection matrix is either $\bm{\Theta}_{0}$ or $\bm{\Theta}_1$.

Before formulating the QoS constraint for the SIR, the detection strategy should be discussed since it is directly related to the QoS performance metric.
To detect the secondary information at the SIR, the authors in \cite{Yan WCL 2020}, \cite{Yan 2019} proposed a two-step approach, which recovers the primary symbols first, and then utilizes the recovered primary symbols to decode the secondary symbols.
On the other hand, the primary and secondary symbols in \cite{Zhang 2020} were jointly recovered using a maximum-likelihood (ML) detector.
However, requiring decoding of the much higher rate primary transmissions places an excessive computational burden on the SIR, which is only interested in the lower rate secondary information.
Therefore, we propose a more efficient strategy to embed the secondary symbols, which facilitates a very simple hard decision detector.

Specially, the detector of SIR simply extracts the binary embedded secondary information by evaluating the average received signal during the $L$ time slots.
To express the average of the SIR's signals collected during the $L$ time slots, we first assume that, in the $l$-th time slot, $l = 1, \ldots, L$, the index of the primary transmitted symbol vector is $m_l \in \left\{1,\ldots,\Omega^K\right\}$.
The signal received by the SIR is thus expressed as
\be
y_{m_l,\text{s},0/1} = (\mathbf{h}_\text{s}^H+\mathbf{h}_{\text{rs}}^H\bm{\Theta}_{0/1}\mathbf{G})\mathbf{x}_{m_l} + n_\text{s},
\ee
where the subscript ``s" denotes the SIR, $\mathbf{h}_\text{s}$ and $\mathbf{h}_{\text{rs}}$ are respectively the channel vectors from the BS to SIR and from the IRS to SIR\footnote{
\textcolor{black}{Although perfect CSI acquisition in IRS-assisted systems is still a challenging task, accurate CSI can be obtained with existing channel estimation algorithms \cite{You 2019}-\cite{Wei TCOM 2021}.
In order to specifically focus on the joint passive reflection and information transmission problems, we assume perfect CSI in this paper.
In addition, the CSI of primary information receivers and secondary information receiver can be distinguished at the BS by different pilot sequences or time division duplexing (TDD) mode.}}, and $n_\text{s}\sim \mathcal{CN}(0,\sigma^2)$ is AWGN at the SIR.
Denoting the set of transmitted symbol indices during the $L$ time slots as $\mathcal{I} \triangleq \left\{m_1,\ldots,m_L\right\}$, the average received signal at the SIR is expressed as
\begin{equation}
\begin{aligned}
\overline{r}_{0/1} &= \frac{1}{L}\sum_{m_l\in \mathcal{I}}y_{m_l,\text{s},0/1} \\
&= \left\{ \begin{array}{l l } \frac{1}{L}(\mathbf{h}_\text{s}^H+\mathbf{h}_{\text{rs}}^H\bm{\Theta}_{0}\mathbf{G})
\sum_{m_l\in \mathcal{I}}\mathbf{x}_{m_l} + n_\text{s} : H_0,\\
  \frac{1}{L}(\mathbf{h}_\text{s}^H+\mathbf{h}_{\text{rs}}^H\bm{\Theta}_{1}\mathbf{G})
\sum_{m_l\in \mathcal{I}}\mathbf{x}_{m_l} + n_\text{s} : H_1.\end{array} \right.
\end{aligned}
\end{equation}
With the principle of binary-PSK (BPSK) modulation, the secondary information is detected as ``1'' if $\mathfrak{R}\{\overline{r}\} > 0$, or ``0'' if $\mathfrak{R}\{\overline{r}\} < 0$.
Inspired by the symbol-level precoding, we utilize the CI region concept for the secondary information transmission design.
The CI regions for the SIR are defined as follows
\begin{subequations}\label{eq:SR QoS}
\begin{align}
&\mathfrak{R}\Big\{\frac{1}{L}(\mathbf{h}_\text{s}^H+\mathbf{h}_{\text{rs}}^H\bm{\Theta}_0\mathbf{G})
\sum_{m_l\in \mathcal{I}}\mathbf{x}_{m_l}\Big\} \leq -\beta : H_0, \\
&\mathfrak{R}\Big\{\frac{1}{L}(\mathbf{h}_\text{s}^H+\mathbf{h}_{\text{rs}}^H\bm{\Theta}_1\mathbf{G})
\sum_{m_l\in \mathcal{I}}\mathbf{x}_{m_l}\Big\} \geq \beta : H_1,
\end{align}
\end{subequations}
where $\beta > 0$ is the QoS requirement that denotes the minimum Euclidean distance between the average received noise-free signal and its decision boundary.
During the $L$ time slots, $\sum_{m_l\in \mathcal{I}}\mathbf{x}_{m_l}$ can take on $C_{\Omega^K}^L$ possible values, which cannot be practically accounted for in the precoding and reflection design.
Thus, we propose to simplify this constraint by decomposing the summation into individual terms.
Then, the QoS constraint for the SIR is re-formulated as
\begin{subequations}\label{eq:SR QoS t}
\begin{align}
&\mathfrak{R}\left\{(\mathbf{h}_\text{s}^H+\mathbf{h}_{\text{rs}}^H\bm{\Theta}_0\mathbf{G})
\mathbf{x}_{m}\right\} \leq -\beta, \forall m, \\
&\mathfrak{R}\left\{(\mathbf{h}_\text{s}^H+\mathbf{h}_{\text{rs}}^H\bm{\Theta}_1\mathbf{G})
\mathbf{x}_{m}\right\} \geq \beta, \forall m,
\end{align}
\end{subequations}
which is a stricter constraint than (\ref{eq:SR QoS}) since the SIR's received noise-free signal is forced to satisfy the QoS requirement in each time slot, rather than the average signal over all $L$ time slots.

Based on the above, the procedure for the IRS-based simultaneous primary and secondary information transmissions is straightforward.
With given QoS requirements of PIRs and SIR, the symbol-level precoders $\mathbf{x}_m$, $\forall m$, and reflection vectors $\bm{\theta}_0$, $\bm{\theta}_1$ are jointly optimized at the BS.
Then, the optimized reflection vectors $\bm{\theta}_0$, $\bm{\theta}_1$ are transmitted to the IRS controller through a dedicated control link.
During the transmission phase, the BS sends the precoded signal $\mathbf{x}_m$ in each time slot according to the corresponding primary information symbol vector $\mathbf{s}_m$.
Meanwhile, the controller adjusts the IRS reflection vector $\bm{\theta}_0$ or $\bm{\theta}_1$ every $L$ time slots according to the binary secondary information.

The design of the precoding vectors $\mathbf{x}_m, \forall m$, and the reflection vectors $\bm{\theta}_0$ and $\bm{\theta}_1$ leads to a joint passive reflection and information transmission problem that is more complicated than that in the previous section.
In the following subsections, we propose efficient algorithms to solve the resulting joint power minimization and QoS balancing optimization problems.

\subsection{Design for Power Minimization}

In this subsection, we investigate the joint design of the precoding and reflection vectors to minimize the transmit power while guaranteeing the QoS for the PIRs (\ref{eq: PR QoS}) and SIR (\ref{eq:SR QoS t}).
Accordingly, the general problem can be formulated as
\begin{subequations}
\label{eq:JAPIT PM}
\begin{align}
\label{eq:JAPIT PM obj}
&\underset{\substack{\mathbf{x}_m, \forall m,\\\bm{\theta}_0, \bm{\theta}_1}}{\min}~~
\sum_{m=1}^{\Omega^K}\left\|\mathbf{x}_m\right\|^2 \\
\label{eq:JAPIT PM c11}
&~\text{s.t.}~~~~~~\mathfrak{R}\left\{\widetilde{y}_{m,k,0/1}\right\}\sin\psi-\left|\mathfrak{I}
\left\{\widetilde{y}_{m,k,0/1}\right\}\right|\cos\psi \geq \alpha_k, \non\\
&\hspace{1.4 cm}\widetilde{y}_{m,k,0/1} = (\mathbf{h}_k^H+\mathbf{h}_{\text{r}k}^H\bm{\Theta}_{0/1}\mathbf{G})
\mathbf{x}_{m}e^{-\jmath\angle s_{m,k}}, \forall m, k, \\
&\hspace{1.4 cm}\mathfrak{R}\left\{(\mathbf{h}_\text{s}^H+\mathbf{h}_{\text{rs}}^H\bm{\Theta}_0\mathbf{G})
\mathbf{x}_{m}\right\} \leq -\beta, \forall m, \\
\label{eq:JAPIT PM c13}
&\hspace{1.4 cm}\mathfrak{R}\left\{(\mathbf{h}_\text{s}^H+\mathbf{h}_{\text{rs}}^H\bm{\Theta}_1\mathbf{G})
\mathbf{x}_{m}\right\} \geq \beta, \forall m, \\
\label{eq:JAPIT PM c2}
&\hspace{1.4 cm}\bm{\Theta}_{0/1} = \text{diag}\{\bm{\theta}_{0/1}\},\;\;\;|\bm{\theta}_{0/1}(n)| = 1, \forall n.
\end{align}
\end{subequations}
Note that this is a non-convex optimization problem that is difficult to solve due to the following reasons.
First, the precoders $\mathbf{x}_m, \forall m$ and the reflection vectors $\bm{\theta}_0$, $\bm{\theta}_1$ are intricately coupled in the QoS constraints (\ref{eq:JAPIT PM c11})-(\ref{eq:JAPIT PM c13}).
Second, the unit-modulus constraint of the reflection vectors in constraint (\ref{eq:JAPIT PM c2}).
To address this problem, we propose to partition all optimization variables properly into two blocks (i.e., $\mathbf{x}_m, \forall m$ and $\bm{\theta}_0$, $\bm{\theta}_1$), and then solve each of the sub-problem alternately until convergence is achieved.

With fixed $\bm{\theta}_0$ and $\bm{\theta}_1$, the compound channel from the BS to the $k$-th PIR and SIR can be concisely expressed as
\begin{equation}\begin{aligned}\label{eq:compound channel}
\widetilde{\mathbf{h}}^H_{k,0/1} &\triangleq \mathbf{h}^H_k + \mathbf{h}^H_{\text{r}k}\bm{\Theta}_{0/1}\mathbf{G}, \forall k, \\
\widetilde{\mathbf{h}}^H_{\text{s},0/1} &\triangleq \mathbf{h}^H_{\text{s}} + \mathbf{h}^H_{\text{rs}}\bm{\Theta}_{0/1}\mathbf{G}.
\end{aligned}
\end{equation}
Similarly, the precoder vectors $\mathbf{x}_m, m = 1, \ldots, \Omega^K$, are independent of each other for the power minimization problem, and thus it can be divided into $\Omega^K$ sub-problems to be solved in parallel.
In particular, the $m$-th sub-problem for optimizing $\mathbf{x}_m$ is given by\vspace{-0.2 cm}
\begin{subequations}
\label{eq:JAPIT PM xm}
\begin{align}
  &\underset{\mathbf{x}_m}{\min}~~\left\|\mathbf{x}_m\right\|^2 \\
  \label{eq:JAPIT PM xm c1}
&~\text{s.t.}~~\mathfrak{R}\left\{\widetilde{\mathbf{h}}^H_{k,0/1}\mathbf{x}_m
e^{-\jmath\angle s_{m,k}}\right\}\sin\psi \non\\ &~~~~~~~~~~-\left|\mathfrak{I}\left\{\widetilde{\mathbf{h}}^H_{k,0/1}\mathbf{x}_m
e^{-\jmath\angle s_{m,k}}\right\}\right|\cos\psi \geq \alpha_k, \forall k, \\
&~~~~~~~\mathfrak{R}\left\{\widetilde{\mathbf{h}}^H_{\text{s},0}\mathbf{x}_{m}\right\} \leq -\beta,  \\
&~~~~~~~\mathfrak{R}\left\{\widetilde{\mathbf{h}}^H_{\text{s},1}\mathbf{x}_{m}\right\} \geq \beta,  \end{align}
\end{subequations}
which is convex and can be solved by standard optimization tools, e.g., CVX.
In addition, the more efficient gradient projection algorithm \cite{CM ITSP 2015} can be employed to solve (\ref{eq:JAPIT PM xm}) and the details are omitted for brevity.

After obtaining precoder vectors $\mathbf{x}_m, m = 1,\ldots,\Omega^K$, the reflection design problem is reduced to a feasibility-check problem without an objective, which may generate many solutions and lead to different convergence rate.
Thus, we utilize an auxiliary variable $t$ to impose stricter QoS constraints, which can provide more freedom for power minimization in the next iteration and accelerate the convergence.
To this end, the IRS reflection design problem is transformed to
\begin{subequations}
\label{eq:JAPIT PM maxmin theta}
\begin{align}
  &\underset{t,\bm{\theta}_0,\bm{\theta}_1}{\max}~~t \\
  &\text{s.t.}~\mathfrak{R}\left\{\widetilde{y}_{m,k,0/1}\right\}\sin\psi-\left|\mathfrak{I}
\left\{\widetilde{y}_{m,k,0/1}\right\}\right|\cos\psi \geq \alpha_kt, \non\\
&~~~~~\widetilde{y}_{m,k,0/1} = (\mathbf{h}_k^H+\mathbf{h}_{\text{r}k}^H\bm{\Theta}_0\mathbf{G})\mathbf{x}_me^{-\jmath\angle s_{m,k}}, \forall m, k, \\
&~~~~-\mathfrak{R}\left\{(\mathbf{h}_\text{s}^H+\mathbf{h}_{\text{rs}}^H\bm{\Theta}_0\mathbf{G})
\mathbf{x}_{m}\right\} \geq \beta t, \forall m, \\
&~~~~\mathfrak{R}\left\{(\mathbf{h}_\text{s}^H+\mathbf{h}_{\text{rs}}^H\bm{\Theta}_1\mathbf{G})
\mathbf{x}_{m}\right\} \geq \beta t, \forall m,\\
&~~~~\bm{\Theta}_{0/1} = \text{diag}\{\bm{\theta}_{0/1}\},\;\;\;|\bm{\theta}_{0/1}(n)| = 1, \forall n,
\end{align}
\end{subequations}
where $t \geq 1$ since the reflection coefficients obtained from the previous iteration satisfy the QoS requirements.
Thus, after solving (\ref{eq:JAPIT PM maxmin theta}), a better QoS than the original requirement is achieved for the obtained precoders in the current iteration.
In order to solve this multivariate problem, we convert it into a univariate problem by eliminating $t$, combing $\bm{\theta}_0$ and $\bm{\theta}_1$ into a single vector, and exploiting the RCG algorithm to handle the unit-modulus constraint, as detailed below.

\begin{algorithm}[!t]\begin{small}
\caption{Joint Symbol-Level Precoding and Reflection Design for the Power Minimization Problem}
\label{alg:PM}
\begin{algorithmic}[1]
\REQUIRE $\mathbf{h}_k$, $\mathbf{h}_{\text{r}k}$, $\alpha_k, \forall k$, $\mathbf{G}$, $\mathbf{h}_\text{s}$, $\mathbf{h}_{\text{rs}}$, $\Omega$, $\sigma^2$, $\beta$.
\ENSURE $\mathbf{X}^\star$, $\bm{\theta}_0^\star$, $\bm{\theta}_1^\star$.
    \STATE {Initialize ${\bm{\theta}} \in \mathcal{M}_\text{cc}$.}
    \STATE{\textbf{Repeat}}
        \STATE{\hspace{0.4 cm}Calculate each precoder vector $\mathbf{x}_m^\star, \forall m$ by solving (\ref{eq:JAPIT PM xm}).}
        \STATE{\hspace{0.4 cm}Obtain continuous $\bm{\theta}^\star$ by solving (\ref{eq:QoS theta}).}
        \STATE{\hspace{0.4 cm}Construct reflection vectors $\bm{\theta}_0^\star$ and $\bm{\theta}_1^\star$ by (\ref{eq:construct theta}).}
        \STATE{\hspace{0.4 cm}Calculate low-resolution solutions using (\ref{eq: theta mn}).}
    \STATE{\textbf{Until convergence}}
    \end{algorithmic}\end{small}
\end{algorithm}

First, in order to re-arrange the optimization problem in (\ref{eq:JAPIT PM maxmin theta}) to a univariate problem, we define
\begin{equation}
\begin{aligned}
\bm{\theta} &\triangleq [\bm{\theta}_0^T, \bm{\theta}_1^T]^T,\;\;\;
a_{m,K+1} \triangleq\mathbf{h}_\text{s}^H\mathbf{x}_m, \forall m, \\
a_{m,k} &\triangleq \mathbf{h}_k^H\mathbf{x}_me^{-\jmath\angle s_{m,k}}, \forall m, k,\\
\mathbf{b}_{m,k,0/1}^H &\triangleq \mathbf{e}_{0/1}^T \otimes  \left\{\mathbf{h}_{\text{r}k}^H \text{diag}\{\mathbf{G}\mathbf{x}_me^{-\jmath\angle s_{m,k}}\}\right\}, \forall m, k, \\
\mathbf{b}_{m,K+1,0/1}^H &\triangleq \mathbf{e}_{0/1}^T \otimes  \left\{\mathbf{h}_{\text{rs}}^H \text{diag}\{\mathbf{G}\mathbf{x}_m\}\right\}, \forall m, \non
\end{aligned}
\end{equation}
where $\mathbf{e}_0 \triangleq [1, 0]^T, \mathbf{e}_1 \triangleq [0, 1]^T$.
Thus, we have the following concise representations of the received signals
\be\begin{aligned}
\widetilde{y}_{m,k,0/1} &= a_{m,k} + \mathbf{b}_{m,k,0/1}^H\bm{\theta}, \forall m, k, \\
(\mathbf{h}_\text{s}^H+\mathbf{h}_{\text{rs}}^H\bm{\Theta}_{0/1}\mathbf{G})
\mathbf{x}_{m} &= a_{m,K+1} + \mathbf{b}_{m,K+1,0/1}^H\bm{\theta}, \forall m.
\end{aligned}\ee
Then, the optimization problem (\ref{eq:JAPIT PM maxmin theta}) is converted to
\begin{subequations}\label{eq:QoS max t}
\begin{align}
&\underset{\bm{\theta},t}{\max}~~~t \\
&~\text{s.t.}~~~f_i \triangleq \mathfrak{R}\{\mathbf{b}_i^H\bm{\theta}\} + w_i \leq -t, \forall i, \\
&~~~~~~~~g_i \triangleq \mathfrak{R}\{\mathbf{c}_i^H\bm{\theta}\} + z_i \leq -t, \forall i, \\
&~~~~~~~~\left|\bm{\theta}(n)\right| = 1, n = 1, \ldots, 2N,
\end{align}
\end{subequations}
where the expressions for $\mathbf{b}_i$ and $\mathbf{c}_i$ are similar to (\ref{eq:bcuv}) and omitted for brevity.
The constant terms $w_i$ and $z_i$, $i = 1, \ldots, 2(K+1)\Omega^K$, are expressed as
\begin{equation}
\begin{aligned}
&w_{2j-1} = w_{2j} = -a_{m,k}\left(\sin\psi - e^{-\jmath\frac{\pi}{2}}\cos\psi\right)/\alpha_k, \\
&w_{2K\Omega^K+m} = a_{m,K+1}/\beta, \\
&z_{2j-1} = z_{2j}= -a_{m,k}\left(\sin\psi + e^{-\jmath\frac{\pi}{2}}\cos\psi\right)/\alpha_k, \\
&z_{2K\Omega^K+m}= -a_{m,K+1}/\beta
,
\end{aligned}
\end{equation}
where $j = K(m-1)+k$.
According to (\ref{eq:QoS max t}b) and (\ref{eq:QoS max t}c), we have $-t \geq \max\{f_i, g_i, \forall i\}$, thus problem (\ref{eq:QoS max t}) is further transformed to the following univariate problem
\begin{subequations}\label{eq:QoS theta}
\begin{align}
\underset{\bm{\theta}}{\min}~~&\underset{i}{\max}~~\{f_i, g_i\} \\
&~\text{s.t.}~~\left|\bm{\theta}(n)\right| = 1, n = 1, \ldots, 2N,
\end{align}
\end{subequations}
which exhibits a similar form as problem (\ref{eq:PIT PM QoS m}).
Therefore, we exploit the log-sum-exp approximation to handle the non-differentiable max value function, and then utilize the RCG algorithm to tackle the non-convex unit-modulus constraint.
The solution to problem (\ref{eq:QoS theta}) follows the same procedures as in Sec. II-B, and thus the details are omitted here for brevity.

After obtaining $\bm{\theta}^\star$, the reflection vectors $\bm{\theta}_0^\star$ and $\bm{\theta}_1^\star$ can be extracted by
\be\label{eq:construct theta}
\bm{\theta}_0^\star = \bm{\theta}^\star(1:N), \;\;\;
\bm{\theta}_1^\star = \bm{\theta}^\star(N+1:2N).
\ee
For the low-resolution cases, we adopt the most efficient direct quantization method as in (\ref{eq: theta mn}).

In summary, given random initial reflection vectors, the precoder vectors $\mathbf{x}_m, \forall m$, and reflection vectors $\bm{\theta}_0, \bm{\theta}_1$, are iteratively updated by solving problems (\ref{eq:JAPIT PM xm}) and (\ref{eq:QoS theta}) until convergence is achieved.
The details of this joint symbol-level precoding and reflection design algorithm for the power minimization problem are summarized in Algorithm 2.
Since the reflection design is suboptimal, the monotonic convergence cannot be theoretically guaranteed.
Alternatively, we provide numerical results in Sec. \ref{sec:simulation results} to show the convergence.

\subsection{Design for QoS Balancing}

In this subsection, we solve the QoS balancing problem for both PIRs and the SIR, which maximizes the minimum weighted QoS for a given average transmit power budget $P$ and is formulated as
\begin{subequations}\label{eq:JAPIT QoS}
\begin{align}
&\underset{\substack{\mathbf{x}_m, \forall m,\\ \bm{\theta}_0,\bm{\theta}_1}}{\max}~~t \\
&\text{s.t.}~~t \leq \rho_k\big[\mathfrak{R}\left\{\widetilde{y}_{m,k,0/1}\right\}\sin\psi - \left|\mathfrak{I}\left\{\widetilde{y}_{m,k,0/1}\right\}\right|\cos\psi\big], \non\\
&\hspace{0.4 cm}\widetilde{y}_{m,k,0/1} = (\mathbf{h}_k^H+\mathbf{h}_{\text{r}k}^H\bm{\Theta}_{0/1}\mathbf{G})\mathbf{x}_me^{-\jmath\angle s_{m,k}}, \forall m, k, \\
&\hspace{0.4 cm}t \leq -\varrho\mathfrak{R}
\left\{(\mathbf{h}_\text{s}^H+\mathbf{h}_{\text{rs}}^H\bm{\Theta}_0\mathbf{G})
\mathbf{x}_{m}\right\}, \forall m,\\
&\hspace{0.4 cm}t \leq \varrho\mathfrak{R}
\left\{(\mathbf{h}_\text{s}^H+\mathbf{h}_{\text{rs}}^H\bm{\Theta}_1\mathbf{G})
\mathbf{x}_{m}\right\}, \forall m,\\
&\hspace{0.4 cm}\bm{\Theta}_{0/1} = \text{diag}\{\bm{\theta}_{0/1}\},\;\;\;|\bm{\theta}_{0/1}(n)| = 1, \forall n, \\
&\hspace{0.4 cm}\sum_{m=1}^{\Omega^K}\left\|\mathbf{x}_m\right\|^2 \leq P\Omega^K,
\end{align}
\end{subequations}
where $\rho_k$ and $\varrho$ are the QoS weights for the $k$-th PIR and SIR, respectively.
As before, we decompose this large-scale optimization problem into several sub-problems and iteratively solve them.

With given reflection vectors $\bm{\theta}_0$ and $\bm{\theta}_1$, the compound channels $\widetilde{\mathbf{h}}^H_{k,0/1}$ and $\widetilde{\mathbf{h}}^H_{\text{s},0/1}$ can be obtained by (\ref{eq:compound channel}), and the precoding design problem is rewritten as
\begin{subequations}\label{eq:JAPIT QoS precoding}
\begin{align}\label{eq:JAPIT QoS precoding obj}
&\underset{\mathbf{x}_m, \forall m, t}{\max}~t \\
&~~\text{s.t.}~~~~t \leq \rho_k\big[\mathfrak{R}\left\{\widetilde{\mathbf{h}}^H_{k,0/1}\mathbf{x}_me^{-\jmath\angle s_{m,k}}\right\}\sin\psi \non\\
&~~~~~~~~~~~~- \left|\mathfrak{I}\left\{\widetilde{\mathbf{h}}^H_{k,0/1}\mathbf{x}_me^{-\jmath\angle s_{m,k}}\right\}\right|\cos\psi\big], \forall m, k, \\
&~~~~~~~~~t \leq -\varrho\mathfrak{R}
\left\{\widetilde{\mathbf{h}}^H_{\text{s},0}
\mathbf{x}_{m}\right\}, \forall m,\\
&~~~~~~~~~t \leq \varrho\mathfrak{R}
\left\{\widetilde{\mathbf{h}}^H_{\text{s},1}
\mathbf{x}_{m}\right\}, \forall m,\\
\label{eq:JAPIT QoS precoding power}
&~~~~~~~~~\sum_{m=1}^{\Omega^K}\left\|\mathbf{x}_m\right\|^2 \leq P\Omega^{K},
\end{align}
\end{subequations}
which is a convex optimization problem.
In order to reduce the computational complexity, we still attempt to decompose it into $\Omega^{K}$ sub-problems and deal with the small-scale $\mathbf{x}_m$ individually.
However, the average power constraint (\ref{eq:JAPIT QoS precoding power}) couples the design of each precoder since the total power must be balanced between them.
Thus, we first explore the relationship between the objective value and the power constraint for each precoder, and then solve the power allocation problem.
In the end, the optimal precoders are obtained based on the obtained allocated power.

Assuming that the transmit power allocated to $\mathbf{x}_m$ is $p_m$, $p_m > 0$ and $\sum_{m = 1}^{\Omega^K}p_m = P\Omega^K$, the $m$-th sub-problem of (\ref{eq:JAPIT QoS precoding}) can be formulated as
\begin{subequations}\label{eq:JAPIT QoS precoding xm}
\begin{align}
&\underset{\mathbf{x}_m, t_m}{\max}~t_m \\
&~\text{s.t.}~~~t_m \leq \rho_k\big[\mathfrak{R}\left\{\widetilde{\mathbf{h}}^H_{k,0/1}\mathbf{x}_me^{-\jmath\angle s_{m,k}}\right\}\sin\psi \non\\
&~~~~~~~~~~~~~~~- \left|\mathfrak{I}\left\{\widetilde{\mathbf{h}}^H_{k,0/1}\mathbf{x}_me^{-\jmath\angle s_{m,k}}\right\}\right|\cos\psi\big], \forall k, \\
&~~~~~~~t_m \leq -\varrho\mathfrak{R}
\left\{\widetilde{\mathbf{h}}^H_{\text{s},0}
\mathbf{x}_{m}\right\},\\
&~~~~~~~t_m \leq \varrho\mathfrak{R}
\left\{\widetilde{\mathbf{h}}^H_{\text{s},1}
\mathbf{x}_{m}\right\}, \\
&~~~~~~\left\|\mathbf{x}_m\right\|^2 \leq p_m,
\end{align}
\end{subequations}
where $t_m$ is the minimum weighted QoS for precoder $\mathbf{x}_m$.
We assume that the optimal solution for this problem is $\mathbf{x}_m^{\star}$ and $t_m^{\star}$.
Notice that the norm of $\mathbf{x}_m^{\star}$ is proportional to $t_m^{\star}$ and $\sqrt{p_m}$, and thus the QoS balancing and power minimization problems will yield the same optimal solutions  within a scaling that depends on the allocated power.
Thus, we can obtain a scaled version of $\mathbf{x}_m^\star$ by solving the power minimization problem (\ref{eq:JAPIT PM xm}) with constraints $\alpha_k = \frac{t_0}{\rho_k}, \forall k$, $\beta = \frac{t_0}{\varrho}$, where $t_0 > 0$ is an arbitrary QoS requirement.
After obtaining the optimal solution $\widetilde{\mathbf{x}}_m^\star $ of   (\ref{eq:JAPIT PM xm}), the optimal solution $\mathbf{x}_m^{\star}$ and $t_m^\star$  of problem (\ref{eq:JAPIT QoS precoding xm}) can be obtained by scaling as
\be\label{eq:xm tm relation}
\mathbf{x}_m^\star = \frac{\sqrt{p_m}\widetilde{\mathbf{x}}_m^\star }
{\left\|\widetilde{\mathbf{x}}_m^\star \right\|},\;\;
t_m^\star = \frac{\sqrt{p_m}t_0}
{\left\|\widetilde{\mathbf{x}}_m^\star \right\|}.
\ee

Then, we need to find the optimal power allocation $p_m$ to balance the QoS requirement $t_m^\star $, which is formulated as
\begin{subequations}
\label{eq:power allocation}
\begin{align}
&\underset{p_m,\forall m, t}{\max}~t \\
&~\text{s.t.}~~~~~~t \leq t_m^\star = \frac{\sqrt{p_m}t_0}
{\left\|\widetilde{\mathbf{x}}_m^\star \right\|}, \forall m, \\
&~~~~~~~~~~\sum_{m=1}^{\Omega^{K}}p_m \leq P\Omega^{K}.
\end{align}
\end{subequations}
Similarly, the optimal solution of this problem can also be obtained by solving the corresponding power minimization problem:
\begin{subequations}\label{eq:power allocation pm}
\begin{align}
&\underset{p_m,\forall m}{\min}~~\sum_{m=1}^{\Omega^K}p_m \\
&~\text{s.t.}~~~~\sqrt{p_m} \geq \frac{\left\|\widetilde{\mathbf{x}}_m^\star \right\|}{t_0}, \forall m,
\end{align}
\end{subequations}
which has the same form as problem (\ref{eq:JAPIT PM xm}) and can be efficiently solved using the projected gradient-based algorithm.
Then, substituting the obtained power allocation $p_m$ into (\ref{eq:xm tm relation}), the precoder for the original QoS balancing problem (\ref{eq:JAPIT QoS}) can be calculated.

In summary, the precoding design problem (\ref{eq:JAPIT QoS precoding}) is solved in three steps: \textit{i)} Solving the power minimization problem (\ref{eq:JAPIT PM xm}) with an arbitrary QoS requirement $t_0$ and obtaining its optimal solution $\widetilde{\mathbf{x}}_m^\star $; \textit{ii)} obtaining the power allocation $p_m, \forall m$, by solving problem (\ref{eq:power allocation pm}); \textit{iii)} calculating the solution $\mathbf{x}_m^\star$ by substituting $p_m$ into (\ref{eq:xm tm relation}).

With fixed precoding vectors $\mathbf{x}_m, \forall m$, the design of the reflection coefficients is similar to problem (\ref{eq:JAPIT PM maxmin theta}), and can be solved using the algorithm proposed in the previous subsection by setting $\alpha_k = \frac{1}{\rho_k}, \forall k$, and $\beta = \frac{1}{\varrho}$.

With the previous developments, the joint symbol-level precoding and reflection design for the QoS balancing problem (\ref{eq:JAPIT QoS}) is straightforward and is summarized in Algorithm 3.
Given random initial reflection vectors, problem (\ref{eq:JAPIT QoS precoding}) and (\ref{eq:JAPIT PM maxmin theta}) are iteratively solved to obtain the precoding vectors $\mathbf{x}_m, \forall m$, and the reflection vectors $\bm{\theta}_0$ and $\bm{\theta}_1$ until convergence is achieved.

\begin{algorithm}[!t]\begin{small}
\caption{Joint Symbol-Level Precoding and Reflection Design for the QoS Balancing Problem}
\label{alg:QoS}
\begin{algorithmic}[1]
\REQUIRE $\mathbf{h}_k$, $\mathbf{h}_{\text{r}k}$, $\rho_k, \forall k$, $\mathbf{G}$, $\mathbf{h}_\text{s}$, $\mathbf{h}_{\text{rs}}$, $\Omega$, $\sigma^2$, $\varrho$, $t_0$.
\ENSURE $\mathbf{X}^\star$, $\bm{\theta}_0^\star$, $\bm{\theta}_1^\star$.
    \STATE {Initialize ${\bm{\theta}} \in \mathcal{M}_\text{cc}$.}
    \STATE{\textbf{Repeat}}
        \STATE{\hspace{0.4 cm}Calculate each precoder vector $\widetilde{\mathbf{x}}_m^\star, \forall m$ by solving (\ref{eq:JAPIT PM xm}).}
        \STATE{\hspace{0.4 cm}Obtain power allocation $p_m, \forall m$ by solving (\ref{eq:power allocation pm}).}
        \STATE{\hspace{0.4 cm}Calculate $\mathbf{x}_m^\star$ by (\ref{eq:xm tm relation}).}
        \STATE{\hspace{0.4 cm}Obtain continuous $\bm{\theta}^\star$ by solving (\ref{eq:QoS theta}).}
        \STATE{\hspace{0.4 cm}Construct reflection vectors $\bm{\theta}_0^\star$ and $\bm{\theta}_1^\star$ by (\ref{eq:construct theta}).}
        \STATE{\hspace{0.4 cm}Calculate low-resolution solutions using (\ref{eq: theta mn}).}
    \STATE{\textbf{Until convergence}}
    \end{algorithmic}\end{small}
\end{algorithm}

\subsection{Computational Complexity Analysis}

For the power minimization problem, the computational complexity using the projected gradient algorithm to solve for $\mathbf{x}_m$ is of order $\mathcal{O}\{M^3\}$.
The complexity required to obtain the reflection vectors is of order $\mathcal{O}\{(2N)^{1.5}\}$.
Thus, the total computational complexity to solve the power minimization problem is of order $\mathcal{O}\{\Omega^K[M^3+(2N)^{1.5}]\}$.
For the QoS balancing problem, the computational complexity of the three steps required to obtain the precoders is of order $\mathcal{O}\{\Omega^KM^3\}$,  $\mathcal{O}\{\Omega^{3K}\}$, and $\mathcal{O}\{M\Omega^{K}\}$, respectively.
Thus, the total computational complexity of the QoS balancing algorithm is of order $\mathcal{O}\{\Omega^K(M^3+\Omega^{2K})+(2N)^{1.5}\}$.

\section{Simulation Results}
\label{sec:simulation results}
\vspace{0.2 cm}

In this section, we provide simulation results to demonstrate the feasibility of IRS-based passive information transmission and illustrate the effectiveness of our proposed algorithms.
We adopt the popular settings in this field as in \cite{Wu TWC 2019}.
The noise power at all receivers is $\sigma^2 = -80$dBm.
The constellation order is $\Omega = 4$, corresponding to QPSK.
The path-loss is modelled as $\text{PL}(d) = C_0\left(d_0/d\right)^\iota$, where $C_0 = -30$dB, $d_0 = 1$m, $d$ is the link distance, and $\iota$ is the path-loss exponent.
\textcolor{black}{The small-scale Rician fading channel is assumed for all links.
For example, the channel from the BS to the IRS $\mathbf{G}$ is modelled as
\be
\mathbf{G} = \sqrt{\frac{\kappa}{\kappa+1}}\mathbf{G}^{\text{LoS}} + \sqrt{\frac{1}{\kappa+1}}\mathbf{G}^{\text{NLoS}},
\ee
where $\kappa$ is the Rician factor set as 3dB, $\mathbf{G}^{\text{LoS}}$ is the line-of-sight (LoS) component determined by the geometric locations of the BS and the IRS, and  $\mathbf{G}^{\text{NLoS}}$ is the non-LoS (NLoS) Rayleigh fading component.}

\subsection{Passive Information Transmission System}

\begin{figure}[!t]
\centering
  \includegraphics[width = 3.1 in]{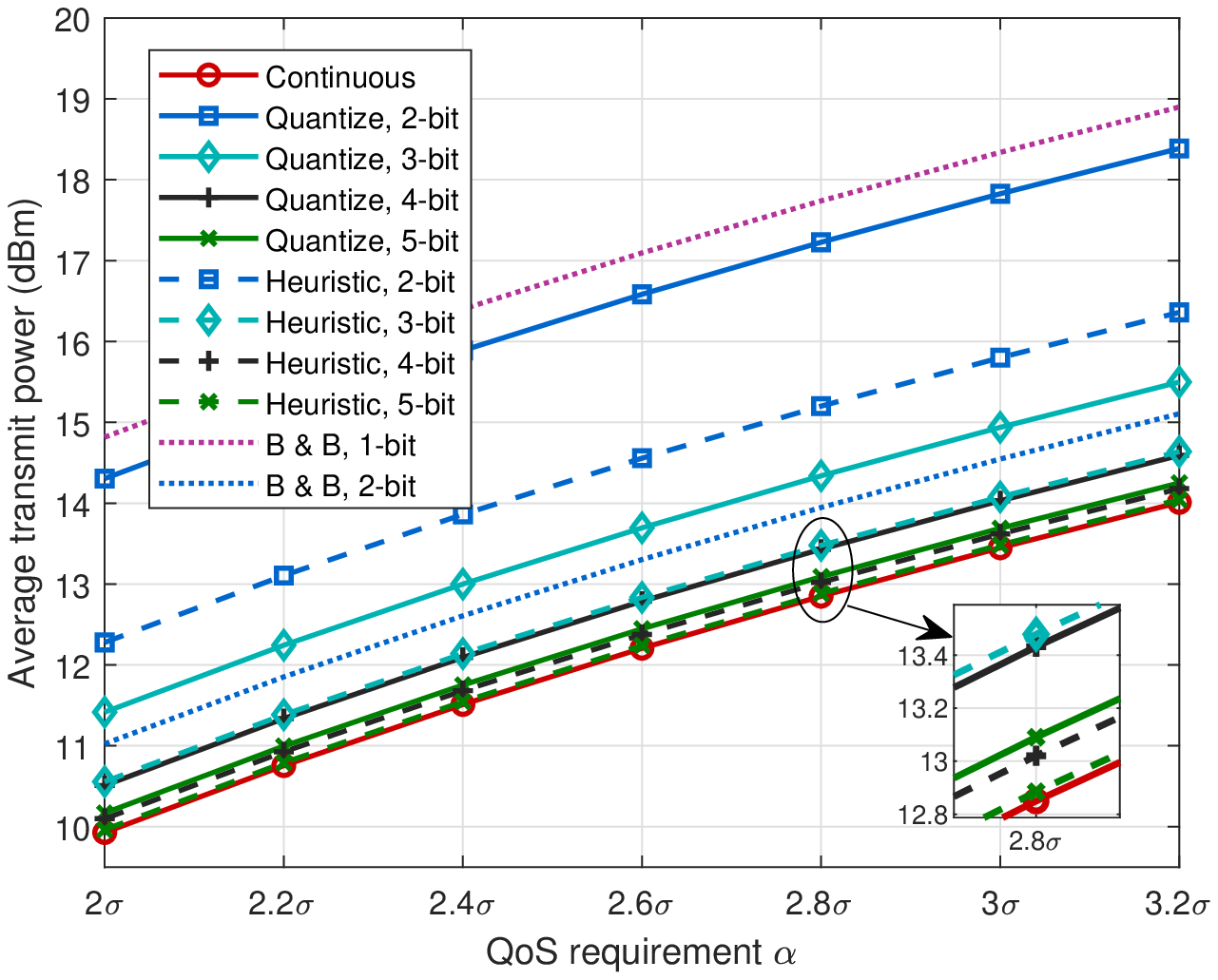}\vspace{0.1 cm}
  \caption{Average transmit power versus QoS requirement $\alpha$ ($K=3$, $N=100$).}
  \label{fig:P_alpha}\vspace{-0.3 cm}
\end{figure}

\begin{figure}[!t]
\centering
\subfigure[Average SER versus $P$.]{
\begin{minipage}{4.2 cm}
\centering
\includegraphics[height = 2.3 in]{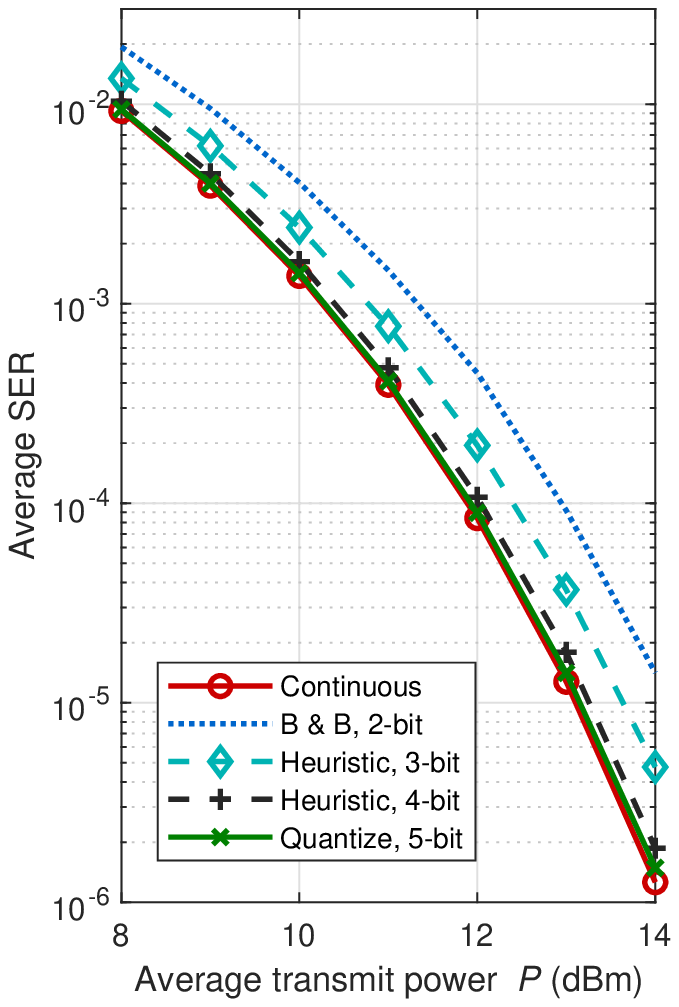}\label{fig:SER_P_av}\vspace{0.1 cm}
\end{minipage}}
\subfigure[Maximum SER versus $P$.]{
\begin{minipage}{4.2 cm}
\centering
\includegraphics[height = 2.3 in]{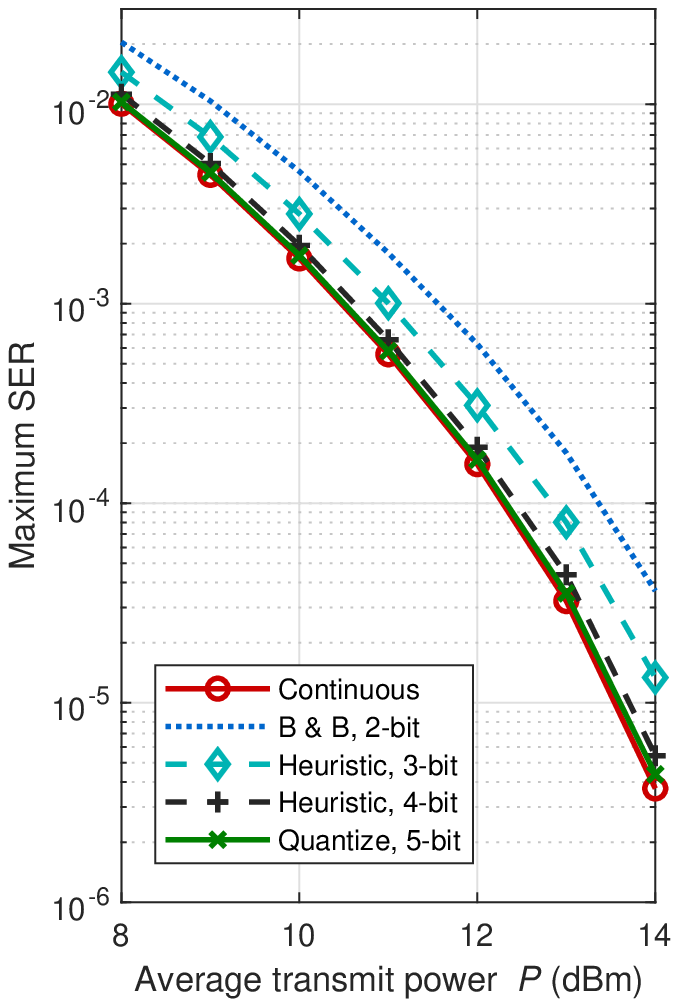}\label{fig:SER_P_m}\vspace{0.1 cm}
\end{minipage}}\vspace{0.1 cm}
\caption{SER versus average transmit power $P$ ($K=3$, $N=100$).}
\label{fig:SER_P}\vspace{-0.3 cm}
\end{figure}

In this subsection, we present simulation results for the passive information transmission system in Sec. II.
We assume that $K=3$ users are randomly distributed 100m away from the IRS, and the path-loss exponent is set as $\kappa = 3$.
For simplicity, we assume that the QoS requirements and weights for all users are the same, i.e., $\alpha_k = \alpha,\;\; \rho_k = \rho = 1,\; \forall k$.

Fig. \ref{fig:P_alpha} shows the average transmit power versus the QoS requirement $\alpha$ for the power minimization problem, including both the continuous and low-resolution cases.
The 2, 3, 4, and 5-bit resolution cases using the direct quantization and heuristic algorithms in Sec. II-D are referred to as ``Quantize, 2-bit''-``Quantize, 5-bit'', and ``Heuristic, 2-bit''-``Heuristic, 5-bit'', respectively.
\textcolor{black}{The 1-bit resolution cases are not shown due to the severely degraded  performance.}
Since the optimal branch-and-bound algorithm has unaffordable exponential complexity for high-resolution cases, only the 1-bit and 2-bit cases are included, which are denoted as ``B \& B, 1-bit'' and ``B \& B, 2-bit'', respectively.
It can be seen that the continuous scheme achieves the best performance given its flexibility in specifying the phase-shifts.
Whereas for a given finite resolution, the branch-and-bound algorithm provides  the best performance and the direct quantization method has the worst performance, as expected given their different levels of computational complexity.
The choice of the algorithm and the resolution of the phase quantization requires a tradeoff between performance and complexity.
For 5-bit or higher resolution, the computational complexity is a more dominant factor in practice, since the performance is already sufficiently close to that with infinite resolution at the IRS.
For the 3-bit and 4-bit cases, the heuristic algorithm is a good choice in the sense that it provides better performance than direct quantization with affordable computational complexity.
For the very low-resolution 1-bit and 2-bit cases, the optimal branch-and-bound algorithm provides much better performance.
However, even using the optimal branch-and-bound algorithm, the 1-bit case still suffers a severe performance loss, requiring almost 5dBm extra power to achieve the same performance as with continuous phase control.
Thus, in Fig. \ref{fig:SER_P} we choose the branch-and-bound algorithm for the 2-bit case, the heuristic algorithm for the 3-bit and 4-bit cases, and the direct quantization method for the 5-bit case to strike a balance between computational complexity and performance.

For the QoS balancing problem, we plot the SER versus the average transmit power $P$ in Fig. \ref{fig:SER_P}.
Since the weight coefficients $\rho_k$ for all receivers are the same, the average and maximum SER of all receivers are plotted in Figs. \ref{fig:SER_P_av} and \ref{fig:SER_P_m}, respectively, to show the QoS balancing performance.
The difference between the average and maximum SER is very small, which verifies the QoS balancing between all receivers.
In addition, it is encouraging to see that the 4-bit and 5-bit schemes achieve almost the same performance as the continuous scheme, and the performance loss of the 2-bit and 3-bit schemes is only 0.5-1dBm.

\begin{figure}[!t]
\centering
\includegraphics[width = 3.1 in]{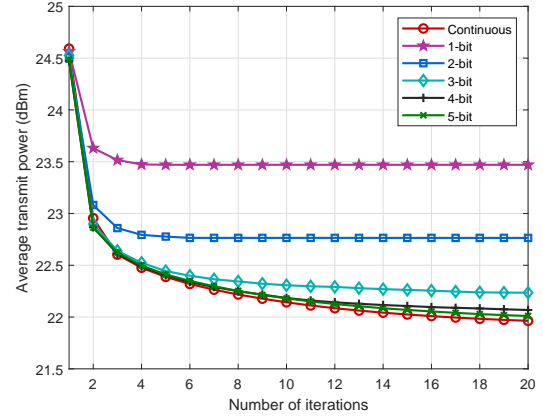}\vspace{0.1 cm}
\caption{Average transmit power versus the number of iterations ($K=3$, $N=100$, $\alpha=2.5\sigma$, $\beta=0.5\sigma$).}\label{fig:P_iter_PM}
\vspace{-0.3 cm}
\end{figure}

\subsection{Joint Passive Reflection and Information Transmission System}

In this subsection, we present simulation results for the joint passive reflection and information transmission designs in Sec. III.
We assume that the BS is equipped with $M = 6$ antennas and serves $K = 3$ PIRs and one SIR.
Since the IRS is usually deployed near the BS or users to achieve more beamforming gains, we assume that the IRS is 10m away from the BS, the PIRs are 100m away from the IRS, and the SIR is 20m away from the IRS to facilitate the secondary information transmission.
Since the BS and IRS are usually deployed at higher elevation to avoid undesired blockages, we assume that the channel between the BS and IRS is stronger than the others.
In particular, the BS-IRS channel is assumed to follow a small-scale Rician fading model with LoS and NLoS components whose path-loss exponent is 2.5, while the other channels only have NLoS components and with a path-loss exponent of 3.
For simplicity, we assume that the QoS requirements and weight coefficients for the PIRs are the same, i.e., $\alpha_k = \alpha$, $\rho_k = \rho = 1$, $\forall k$, and the weight coefficient for the SIR is $\varrho = 5$.

\begin{figure}[!t]
\centering
\includegraphics[width = 3.1 in]{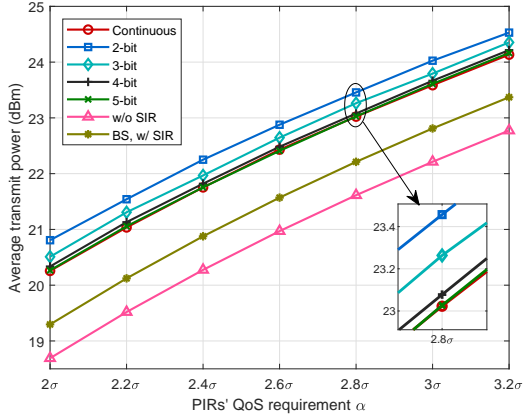}\vspace{0.1 cm}
\caption{Average transmit power versus PIRs' QoS requirement $\alpha$ ($K=3$, $N=100$, $\beta = 0.5\sigma$).}\label{fig:P_alpha_PM}
\vspace{-0.3 cm}
\end{figure}

\begin{figure}[!t]
\centering
  \includegraphics[width = 3.1 in]{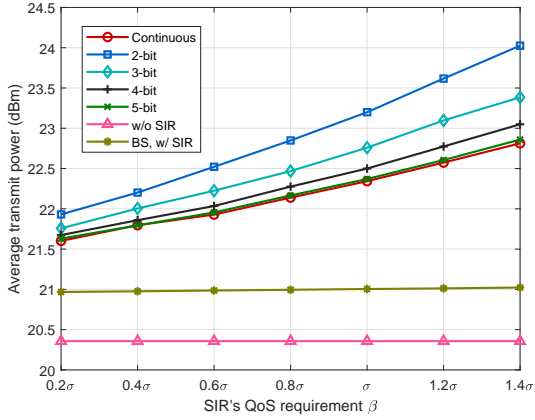}\vspace{0.1 cm}
  \caption{Average transmit power versus SIR's QoS requirement $\beta$ ($K=3$, $N=100$, $\alpha = 2.5\sigma$).}  \label{fig:P_beta_PM}
  \vspace{-0.3 cm}
\end{figure}

The simulation results for the power minimization problem are shown in Figs. \ref{fig:P_iter_PM}-\ref{fig:P_beta_PM}.
We first demonstrate the convergence performance in Fig. \ref{fig:P_iter_PM}, where the average transmit power versus the number of iterations is plotted.
It shows that the convergence can be achieved within 20 iterations, which demonstrates a reasonable computational complexity.
Furthermore, we observe that the low-resolution cases have faster convergence than the continuous counterpart.
As in Fig. 4, 1-bit phase-shifts show a notable performance loss due to the quantization error and thus will not be evaluated in the rest of the simulation studies.

Fig. \ref{fig:P_alpha_PM} shows the average transmit power versus the PIR QoS requirement $\alpha$ with fixed SIR QoS requirement $\beta = 0.5\sigma$.
For comparison, we also show the performance for the case where there is no SIR and the IRS only works as a passive reflector for enhancing the primary information transmissions.
These results are plotted as a benchmark and denoted by ``w/o SIR''.
We also consider the scheme where the IRS transfers the secondary information to the BS and the BS simultaneously serves the PIRs and SIR with the aid of the IRS.
This case is denoted as ``BS, w/ SIR''.
The ``w/o SIR'' scheme naturally consumes the least power since it only serves the PIRs.
The ``BS, w/ SIR'' scheme also consumes less power than the proposed schemes due to the BS's powerful processing ability.
However, for this case, the IRS must transfer secondary information to the BS, which requires additional power consumption at the IRS and a higher transmission bandwidth for the control link between the BS and IRS.
Moreover, compared with the case where there is no SIR, less than 1.5dBm extra power is required, which shows the effectiveness of the proposed scheme.
Since the symbol-level precoding at the BS provides significant additional DoFs, the performance loss due to quantization of the IRS phases is much lower in this scenario compared with that in Fig. 4.

\begin{figure}[!t]
\centering
  \includegraphics[width = 3.1 in]{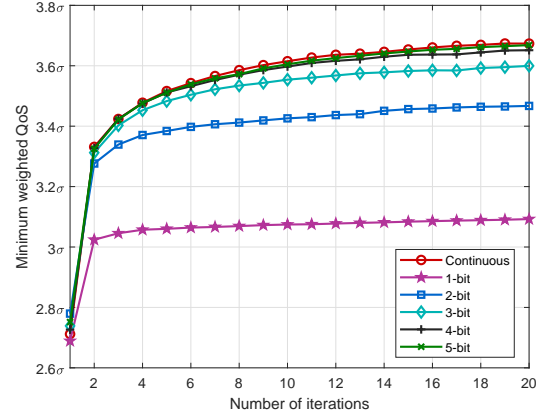}\vspace{0.1 cm}
  \caption{Minimum weighted QoS versus the number of iterations ($K=3$, $N=100$, $P = 25$dBm).}\label{fig:t_iter_QoS}
  \vspace{-0.3 cm}
\end{figure}

\begin{figure}[!t]
\centering
\subfigure[Average SER versus $P$.]{
\begin{minipage}{4.2 cm}
\centering
\includegraphics[height = 2.3 in]{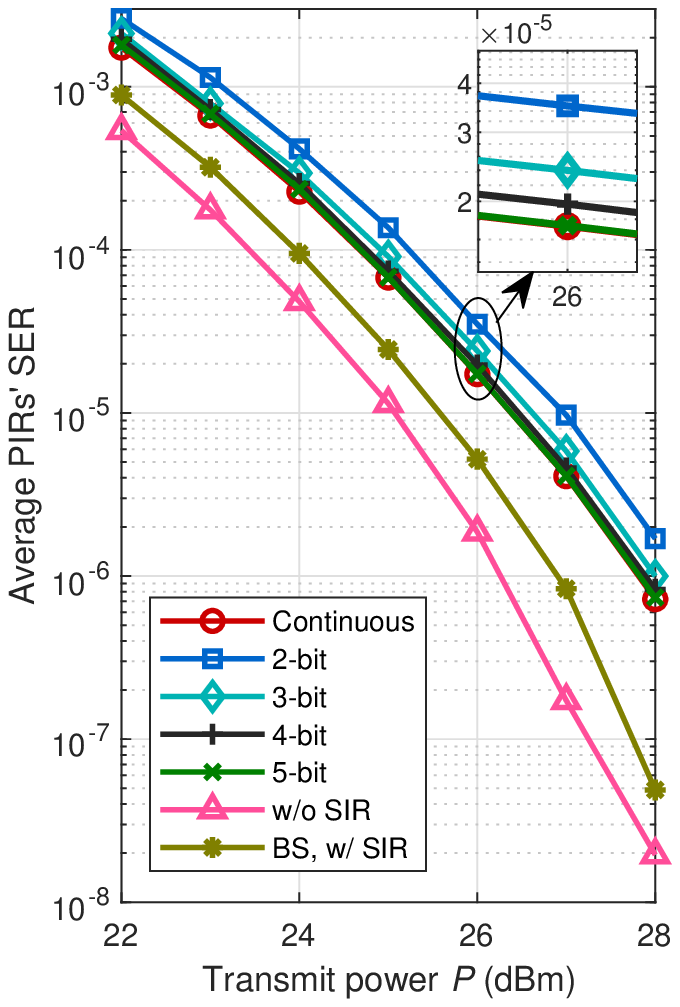}\label{fig:SER_P_av_QoS}\vspace{0.1 cm}
\end{minipage}}
\subfigure[Maximum SER versus $P$.]{
\begin{minipage}{4.2 cm}
\centering
\includegraphics[height = 2.3 in]{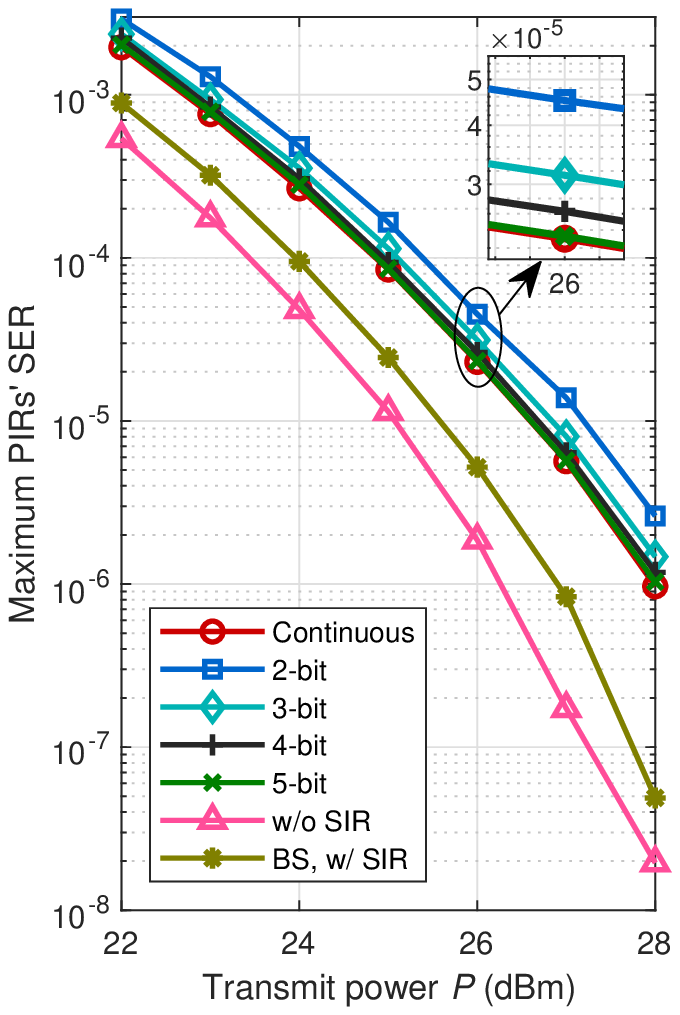}\label{fig:SER_P_m_QoS}\vspace{0.1 cm}
\end{minipage}}\vspace{0.1 cm}
\caption{SER versus average transmit power $P$ ($K=3$, $N=100$).}
\label{fig:SER_P_QoS}
\vspace{-0.3 cm}
\end{figure}

In Fig. \ref{fig:P_beta_PM}, we plot the average transmit power versus the SIR QoS requirement $\beta$ with fixed PIR QoS requirement $\alpha = 2.5\sigma$. We see that as $\beta$ increases, only a small amount of extra power is required to provide better QoS for the SIR, especially for the high-resolution cases, which illustrates the effectiveness of our proposed scheme for embedding the secondary symbols into primary transmissions.

\begin{figure}[!t]
\centering
  \includegraphics[width = 3.1 in]{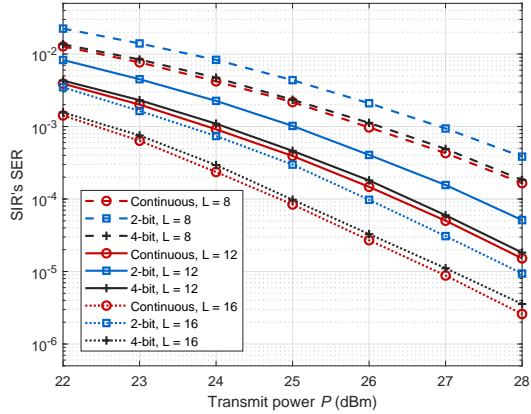}\vspace{0.1 cm}
  \caption{Average SER of SIR versus average transmit power $P$ ($K=3$, $N=100$).}
  \label{fig:SR_SER_P}
  \vspace{-0.3 cm}
\end{figure}

Figs. \ref{fig:t_iter_QoS}-\ref{fig:SR_SER_P} show simulation results for the QoS balancing problem.
In Fig. \ref{fig:t_iter_QoS}, the minimum weighted QoS versus the number of iterations is plotted to show the convergence of the proposed algorithm.
As in the previous examples, convergence is faster for the low-resolution cases, but in all cases it is achieved relatively quickly, within about 20 iterations, although little performance improvement is observed beyond about 10 iterations.

In Fig. \ref{fig:SER_P_QoS}, the average and maximum PIR SER versus transmit power $P$ are plotted to show the QoS balancing performance.
We observe that the performance gap between Fig. \ref{fig:SER_P_av_QoS} and Fig. \ref{fig:SER_P_m_QoS} is relatively small, which fits well with the setting $\rho_k = 1, \forall k$.
In addition, comparing the ``BS, w/ SIR'' and ``w/o SIR'' schemes, the performance loss of the proposed continuous approach is about 1dBm and 1.5dBm, respectively, which is a reasonable cost for transmitting the additional secondary information.
Moreover, the efficient low-resolution schemes also have encouraging performance.

The SIR SER versus transmit power $P$ is shown in Fig. \ref{fig:SR_SER_P}, including continuous, 2-bit, and 3-bit phase resolution with different embedding rates, $L = 8, 12, 16$.
We observe that SER decreases with increasing $L$, revealing the trade-off between efficiency and reliability.

\section{Conclusions}\label{sec:conclusion}
\vspace{0.2 cm}

In this paper, we investigated IRS-assisted passive information transmissions in downlink MU-MISO systems by exploiting the symbol-level precoding.
A dedicated passive information transmission system was first considered, where the IRS operates as a passive transmitter by reflecting an unmodulated carrier signal from an RF generator to multiple users.
Then, a joint passive reflection and information transmission system was investigated, where the IRS enhances primary information transmissions and simultaneously delivers its own secondary information.
Efficient algorithms were proposed to solve the power minimization and QoS balancing problems for both systems.
Extensive simulation results confirmed the feasibility of IRS-assisted passive information transmission and the effectiveness of the proposed algorithms.

\end{document}